\documentclass{article}
\usepackage[utf8]{inputenc}

\usepackage{amsmath}

\usepackage{amssymb}

\newtheorem{theorem}{Theorem}
\newtheorem{corollary}[theorem]{Corollary}

\newtheorem{lemma}[theorem]{Lemma}
\newtheorem{claim}[theorem]{Claim}
\newtheorem{definition}[theorem]{Definition}
\newtheorem{proposition}[theorem]{Proposition}

\newtheorem{remark}[theorem]{Remark}

\newenvironment{proof}{\noindent\bf{Proof.}\rm}{\hfill$\blacksquare$\bigskip}

\newcommand{\items}{\mathcal{M}} 
\newcommand{\agents}{\mathcal{N}} 

\begin{document}

\title{Fair allocations with subadditive and XOS valuations}

\author{Uriel Feige\thanks{Weizmann Institute, Israel. {\tt uriel.feige@weizmann.ac.il}} \and Vadim Grinberg\thanks{Weizmann Institute, Israel. {\tt vadim.grinberg@weizmann.ac.il}}}

\maketitle

\begin{abstract}
  We consider the problem of fair allocation of $m$ indivisible goods to $n$ agents with either subadditive or XOS valuations, in the arbitrary entitlement case. As fairness notions, we consider the anyprice share (APS) ex-post, and the maximum expectation share (MES) ex-ante. 
  
  We observe that there are randomized allocations that ex-ante are at least $\frac{1}{2}$-MES in the subadditive case and $(1-\frac{1}{e})$-MES in the XOS case. Our more difficult results concern ex-post guarantees. We show that $(1 - o(1))\frac{\log\log m}{\log m}$-APS allocations exist in the subadditive case, and $\frac{1}{6}$-APS allocations exist in the XOS case. For the special case of equal entitlements, we show $\frac{4}{17}$-APS allocations for XOS.

  Our results are the first for subadditive and XOS valuations in the arbitrary entitlement case, and also improve over the previous best results for the equal entitlement case.
\end{abstract}

\section{Introduction}

We consider the problem of fair allocation of a set $\items$ of $m$ indivisible goods to $n$ agents. Every agent $i$ has a valuation function $v_i$ over the goods, where $v_i$ is {\em normalized} (the empty set has value~0) and monotone non-decreasing ($v_i(S) \le v_i(T)$ if $S \subset T \subseteq \items$). Agents may have different entitlements to the goods. The entitlement $b_i$ of each agent $i$ is positive, with $\sum_i b_i = 1$. We wish the allocation to be fair, in the sense that it (approximately) satisfies some fairness benchmark. The fairness notion that we use is the anyprice share (APS)~\cite{BEF21APS}. (For definitions of fairness notions that we consider, see Section~\ref{sec:fairness}.) A $\rho$-APS allocation is one that gives each agent a bundle of value at least a $\rho$ fraction of her APS.

It was previously shown that if agents have additive valuations (for definitions of classes of valuations, see Section~\ref{sec:valuations}), then there are $\frac{3}{5}$-APS allocations~\cite{BEF21APS}, and that if agents have submodular valuations, then there are $\frac{1}{3}$-APS allocations~\cite{BUF23}. Moreover, both results were obtained by designing strategies for the {\em bidding game}, which is an allocation mechanism proposed in~\cite{BEF21APS}. (See Section~\ref{sec:bidding} for more details on the bidding game.)

In this work we design approximate APS allocation algorithms for classes of valuations beyond submodular. Specifically, we consider XOS valuations and subadditive valuations. (Note that each of the classes additive, submodular, XOS and subadditive is strictly more general than the one preceding it.) This seems challenging. For example, as shown in~\cite{BUF23}, for the plain version of the bidding game, no strategy can guarantee more than $O(m^{-1/3)}$-APS when valuations are XOS. 
%whereas there are strategies that offer $\Omega(1)$-APS when valuations are submodular. 
So far, no techniques other than the bidding game have been used in order to obtain approximate APS allocations, when agents have arbitrary entitlements.

What can we hope to achieve? It is known that even in the easier case of equal entitlements and even with respect to the easier benchmark of the maximin share (MMS), there are instances with XOS valuations in which there are no allocations better than $\frac{1}{2}$-MMS~\cite{GhodsiHSSY22}.  Hence the best that we can hope for is $\frac{1}{2}$-APS allocations, and there is no example that we are aware of that shows that this cannot always be achieved for subadditive valuations. (For more general classes of monotone valuations, there are simple examples showing that no approximation of the MMS is possible.)

An encouraging sign in this direction is the case of ex-ante share based guarantees. Here a natural share is the maximum expected share (MES), and one wishes to find a distribution over allocations such that for every agent, in expectation she receives a bundle whose value is a handsome fraction of her MES. For subadditive valuations, $\frac{1}{2}$-MES randomized allocations exist, and for XOS valuation, even $(1 - \frac{1}{e})$-MES randomized allocations exist (where $1 - \frac{1}{e} \simeq 0.632$). See Theorem~\ref{thm:exante}.

Our main result for subadditive valuations is the existence of $(1 - o(1))\frac{\log\log m}{\log m}$-APS allocations. Beyond handling the case of arbitrary entitlements, this also improves over the best previous ratios in the simpler equal entitlements case. Interestingly, this result is obtained by designing a strategy for the bidding game, but not for its plain version (for which such a result would be impossible, by~\cite{BUF23}), but for its extended version. This is the first provable separation between the guarantees obtainable between the two versions of the bidding game, and suggests that if the bidding game is to be used as an allocation mechanism in practice, it is preferable to use its extended version.

For XOS valuations, it was previously known that with equal entitlements, there are $\frac{3}{13}$-MMS allocations~\cite{GhodsiHSSY22,SS24,AkramiMSS23}. In contrast, we show that in the bidding game (even in its extended version), there is no strategy that in the equal entitlements case ensures more than $(1 + o(1))\frac{\log\log m}{\log m}$-MMS. 
%Hence, if we wish to extend the $\Omega(1)$ approximation to the arbitrary entitlements case (and APS), there is need for allocation algorithms for the arbitrary entitlement case that are not based on the bidding game. No such algorithms were previously designed.

Our main result for XOS valuations is the existence of $\frac{1}{6}$-APS allocations (in the arbitrary entitlement case). Our algorithm is inspired by the algorithm of~\cite{GhodsiHSSY22}. However, a preliminary step of that algorithm (which allows one to reach a situation in which {for every agent}, no single item has high value) uses in an essential way the assumption of equal entitlements, and fails in the arbitrary entitlements case. To avoid the use of this preliminary step, we modify the allocation algorithm, and introduce new arguments in its analysis.

In the equal entitlement case, our $\frac{1}{6}$-APS result is incomparable with the known $\frac{3}{13}$-MMS result (our approximation ratio is worse, but relative to a more demanding benchmark). To make our improvements more conclusive, we also design a different algorithm (a natural extension of the algorithm of~\cite{AkramiMSS23}) that in the equal entitlement case gives $\frac{4}{17}$-APS allocations (note that the approximation is with respect to the APS, not just the MMS).

\subsection{Classes of valuation functions}
\label{sec:valuations}

Given a set $\items$ of items that are {\em goods} (items that agents want to receive), a valuation function $v : 2^{\items} \rightarrow \Re$ is a set function that is normalized ($v(\emptyset) = 0$) and monotone ($v(S) \le v(T)$ for $S \subset T$). We shall consider classes of valuations that belong to the {\em complement free} hierarchy of valuations, introduced in~\cite{LLN}.

\begin{itemize}

\item Additive. $v(S) = \sum_{e \in S} v(e)$ for every set $S \subset \items$. 

\item Submodular. $v(S \cup \{e\}) - v(S) \ge v(T \cup \{e\}) - v(T)$ for every item $e \in \items$ and sets $S \subset T \subset \items$.

\item XOS. There are additive valuations $v_1, \ldots, v_t$ (for some $t$), and $v(S) = \max_{j \in \{1, \ldots, t\}} v_j(S)$ for every set $S \subset \items$. (An alternative definition of XOS, as fractionally subadditive valuations, appears in~\cite{Feige09}.)

\item Subadditive. $v(S) + v(T) \ge v(S \cup T)$ for all sets $S,T \subset \items$.

\end{itemize}

It is well known that each of the above classes is strictly contained in the class that follows it.

\subsection{Share based fairness notions}
\label{sec:fairness}

In this work, we shall consider three share based fairness notions. For ex-post fairness we consider the maximin share (MMS)~\cite{Budish11} which applies to allocation instances with equal {entitlements}, and
the anyprice share (APS)~\cite{BEF21APS}  which applies to allocation instances with {arbitrary (possibly unequal) entitlements}. For ex-ante fairness we consider the maximum expectation share (MES). A variation on this share was previously defined for the equal entitlements case in~\cite{BF24}, whereas our definition applies also to the arbitrary entitlements case.

\begin{definition}
    For a set $\items$ of items and integer $n \ge 2$, an $n$-partition $(P_1, \ldots, P_n)$ is a partition {of} $\items$ into $n$ disjoint sets. The set of all $n$ partitions is denoted by ${\mathcal{P}}_n$.

    Given a fraction $\rho$, a fractional $\rho$ partition is a collection of bundles $\{P_j\}$ (where $j$ ranges over some index set) and associated nonnegative weights $\{\lambda_j\}$ satisfying:

    \begin{itemize}
        \item $\sum_j \lambda_j = 1$.
        \item $\sum_{j \; \mid \; e \in P_j} \lambda_j = \rho$ for every item $e\in \items$.
    \end{itemize}

    The set of all fractional $\rho$ partitions is denoted by ${\mathcal{FP}}_{\rho}$.
\end{definition}

Observe that every $n$-partition is also a fractional $\frac{1}{n}$ partition.

\begin{definition}
    For an agent $i$ with valuation $v_i$ and entitlement $b_i = \frac{1}{n}$, her maximin share (MMS) is:

    $$MMS(\items,v_i,\frac{1}{n}) = \max_{(P_1, \ldots, P_n) \in {\mathcal{P}}_n} \min_j v_i(P_j)$$
\end{definition}

\begin{definition}
    For an agent $i$ with valuation $v_i$ and entitlement $b_i$, her anyprice share (APS) is:

    $$APS(\items,v_i,b_i) = \max_{(P_1, P_2, \ldots) \in {\mathcal{FP}}_{b_i}} \min_j v_i(P_j)$$
\end{definition}

We note that there is also an equivalent price based definition for the APS that we shall not need to use in this work. See~\cite{BEF21APS}.

For entitlements of the form $\frac{1}{n}$ (for which the MMS is defined), $APS \ge MMS$.  For subadditive valuations, an example in appendix C of the arxiv version of~\cite{BEF21APS} shows a valuation function such that with entitlement $\frac{1}{2}$, the APS is twice as large as the MMS. In Proposition~\ref{pro:APSMMSSubadditive} we prove that for subadditive valuations, the APS is at most five times as large as the MMS. For XOS, our Theorem~\ref{thm:APSXOS4} implies that the APS is at most $\frac{17}{4}$ as large as the MMS. For general monotone valuations, it may happen that the MMS is~0 whereas the APS is positive.

\begin{definition}
    For an agent $i$ with valuation $v_i$ and entitlement $b_i$, her maximum expectation share (MES) is:

    $$MES(\items,v_i,b_i) = \max_{(P_1, P_2, \ldots) \in {\mathcal{FP}}_{b_i}} E_{j}[v_i(P_j)]$$

    The expectation $E_j$ is computed over bundles $B_j$ selected with probability $\lambda_j$, as given by the underlying fractional $b_i$ partition.
\end{definition}

It follows by definition that $MES \ge APS$. It may happen that the MES is positive but the APS is~0 (for example, if there is only one item). We remark that for additive valuations, the MES equals the proportional share, $b_i \cdot v_i(\items)$.

%Like approximate APS allocations for other classes of valuations (additive~\cite{BEF22} and submodular~\cite{BF23}), our proof of Theorem~\ref{thm:APSmain} is based on designing a {\em safe strategy} for the {\em bidding game} of~\cite{BEF22}. Specifically, our result is derived for the version of the game in which the winner of a round can select multiple items (and pay her bid for each selected item). In contrast, it was previously shown in~\cite{BF23} that if the winner is allowed to select only one item, no strategy can guarantee more than $\frac{1}{m^{1/3}}$-APS. Hence our result demonstrates an exponential separation between the guarantees that one can obtain in the two versions of the bidding game, whereas previously no separation was proved. 

%A less ambitious conjecture is that $\rho_{MMS}(n,m) = \frac{1}{2}$ in the equal entitlements case. We confirm that this indeed holds in two special cases.

%\begin{theorem}
%\label{thm:MMSmain}
%    For agents with equal entitlements and subadditive valuations, there always is an allocation that gives each agent at least half of her MMS  (that is, $\rho_{MMS}(n,m) = \Omega(\frac{1}{2})$) in the following special cases:
%\begin{enumerate}
%    \item There are at most~3 agents.
%    \item There are only two different valuation functions (and any number of agents).
%\end{enumerate}
%\end{theorem}

\subsection{Our results}

Our positive results concern agents with arbitrary entitlements (unless stated otherwise), whereas our negative results hold also for agents with equal entitlements. 

The first result that we present characterizes the best approximation ratios with respect to the MES that can be guaranteed ex-ante, both for the case of subadditive valuations and for the case of XOS valuations. Recall that $MES \ge APS$, and that for equal entitlements, $APS \ge MMS$. The negative examples showing tightness of the results hold even in the equal entitlements case and with respect to the MMS. 

\begin{theorem}
    \label{thm:exante}
    The following ex-ante guarantees hold for randomized allocations. 
    \begin{enumerate}
        \item For subadditive valuations:
        \begin{enumerate}
            \item For agents with arbitrary entitlements there is a randomized allocation that gives every agent at least $\frac{1}{2} MES$ in expectation. 
            \item For agents with equal entitlements and every $\epsilon > 0$ there are instances in which no randomized allocation gives every agent more than $(\frac{1}{2} + \epsilon)  MMS$ in expectation. 
        \end{enumerate}
        \item For XOS valuations:
        \begin{enumerate}
            \item For agents with arbitrary entitlements there is a randomized allocation that gives every agent at least $(1 - \frac{1}{e}) MES$ in expectation. 
            \item For agents with equal entitlements and every $\epsilon > 0$ there are instances in which no randomized allocation gives every agent more than $(1 - \frac{1}{e} + \epsilon)  MMS$ in expectation. 
        \end{enumerate}
    \end{enumerate}    
\end{theorem}

The proof of Theorem~\ref{thm:exante} follows easily from the results of~\cite{Feige09}. %We included Theorem~\ref{thm:exante} here mainly because this result seems to have been overlooked in previous work on ex-ante guarantees.

A central result of our paper provides ex-post guarantees for agents with subadditive valuations. This is achieved by designing a {\em safe strategy} for the bidding game of~\cite{BEF21APS}.

\begin{theorem}
\label{thm:biddingStrategy}
    {Let $k$ be the smallest integer such that the number of items satisfies $m \le (k-1)^{k-1}$. For an agent with a subadditive valuation, there is a strategy in the bidding game that ensures the agent at least a $\frac{1}{k}$ fraction of her APS.} 
    Consequently, for agents with arbitrary entitlements and subadditive valuations,  there always are $\frac{(1 - o(1))\log\log m}{\log m}$-APS allocations.
\end{theorem}

The strategy referred to in Theorem~\ref{thm:biddingStrategy} is simple, but its analysis is not. Theorem~\ref{thm:biddingStrategy} has the following consequences.

\begin{itemize}     
    \item For reasonable values of $m$, the approximation ratios offered by Theorem~\ref{thm:biddingStrategy} are moderate size constants. For example, for $m = 3000$ the approximation ratio is $\frac{1}{6}$, whereas for $m = 300000000$ the approximation ratio is $\frac{1}{10}$.
    \item The bidding game comes in two variations, the {\em plain} version and the {\em extended} version (see Section~\ref{sec:bidding}). Theorem~\ref{thm:biddingStrategy} refers to the extended version, whereas for the plain version it is known that even for XOS valuations, there is no strategy that guarantees more than an $O({m^{-1/3}})$ of the APS~\cite{BUF23}. Hence our result shows an exponential separation between the guarantees of these two versions, and strongly suggests that if the bidding game is to be used as an allocation procedure in practice, then one should use the extended version.
    \item The approximation ratio in Theorem~\ref{thm:biddingStrategy} is achieved by a strategy in the bidding game. Strategies for agents with additive valuations are presented in~\cite{BEF21APS}, and for submodular valuations are presented  in~\cite{BUF23}. By each agent picking the strategy designed for her valuation function, this implies that in every allocation instance with arbitrary entitlements there is an allocation that simultaneously gives those agents with subadditive valuations at least $(1 - o(1))\frac{\log\log m}{\log m}$ -APS, those agents with submodular valuations at least $\frac{1}{3}$-APS, and those agents with additive valuations at least $\frac{3}{5}$-APS.
    \item Even in the equal entitlements case, the best approximate MMS guarantee previously known as a function of $m$ was $\Omega(\frac{1}{\log m})$~\cite{GhodsiHSSY22}. Beyond handling the arbitrary entitlement case, Theorem~\ref{thm:biddingStrategy} improves over this in several respects. It shaves off a $\log\log m$ factor, improves the leading constant to a constant that tends to~1 as $m$ grows,  and provides approximation with respect to the more demanding APS, rather than the MMS.
    \item {In the equal entitlements case, Theorem~\ref{thm:biddingStrategy} implies that $\frac{1}{n+1}$-MMS allocations exist, because for the purpose of MMS allocations, one can always assume that $m \le n^n$ (see~\cite{SS24}, for example). In terms of asymptotic dependence on $n$, this is not as good as the $\frac{1}{14 \log n}$-MMS bound of~\cite{FH25}. However, our result has no hidden constants, 
    and so our result {gives better approximation ratios} for a wide range of values of $n$.}
\end{itemize}

The strategy in the proof of Theorem~\ref{thm:biddingStrategy} offers the best possible worst case guarantee (up to low order terms) in the bidding game, even compared to easier cases in which valuations are XOS, agents have equal entitlements and the approximation is with respect to the MMS (instead of the APS).

\begin{theorem}
\label{thm:negative}
    In the bidding game, for every $\epsilon > 0$, there are input instances with $m$ items and $n$ agents with equal entitlements in which an agent $i$ has an XOS valuation, but has no strategy that guarantees more than 
    $$(1 + \epsilon)\frac{\log\log m}{\log m} \cdot MMS(\items, v_i, \frac{1}{n}).$$ 
    {Moreover, in this class of instances the approximation ratio is no better than $O(\frac{1}{\sqrt{n}})$.}
\end{theorem}

For XOS valuations, we show that constant approximations of the APS are possible, despite the fact that they cannot be guaranteed by the bidding game.

\begin{theorem}
\label{thm:APSXOS6}
    In every allocation instance in which every agent $i$ has an XOS valuation $v_i$ and entitlement $b_i$ (with $\sum_{i=1}^n b_i = 1$), there is an allocation in which every agent $i$ gets a bundle of value at least $\frac{1}{6}APS(\items,v_i,b_i)$. 
\end{theorem}

The proof of Theorem~\ref{thm:APSXOS6} draws some inspiration from a proof given in~\cite{GhodsiHSSY22} that in the equal entitlement case, a $\frac{1}{5}$-MMS allocation exists. That proof showed that after a certain reduction step (giving large valued items to agents that are satisfied by them), any allocation that maximizes welfare with respect to a truncated version of the valuations gives the desired guarantees. However, being in the arbitrary entitlement case, the reduction step cannot be applied (technically, the problem is that giving a highly valued item to an agent of low entitlement will reduce the APS of the remaining agents). Without the reduction step, the proof requires new ideas. We prove that starting from any allocation, there is a sequence of allocations, where each one differs from the previous one by replacing the bundle of one agent, such that the final allocation in the sequence satisfies the theorem. It may happen that some steps in the sequence reduce welfare, but nevertheless, we use a welfare based argument to prove that the sequence must terminate.

Finally, equipped with insights obtained by considering the APS rather than the MMS, we revisit the equal entitlement case with XOS valuations.  We improve over the previously best ratio of $\frac{3}{13}$~\cite{AkramiMSS23}, and do so even with respect to the APS instead of the MMS.

\begin{theorem}
\label{thm:APSXOS4}
    In every allocation instance in which every agent $i$ has an XOS valuation $v_i$ and the $n$ agents have equal entitlements, there is an allocation in which every agent $i$ gets a bundle of value at least $\frac{4}{17}APS(\items,v_i,\frac{1}{n})$. 
\end{theorem}

Computational issues related to our allocation algorithms depend on the representations of the underlying valuation functions, or on the kind of queries that can be answered about them efficiently. A short discussion of the types of queries that suffice in order to perform some of the computations in our algorithms is presented in Section~\ref{sec:computation}.

\subsection{Related work}

Let us start by noting that there is extensive literature on aspects of fair allocations that are not considered in our work. We mention some of these aspects here. Our work considers allocation of indivisible goods. There is also much work on allocation of divisible items (sometimes referred to as {\em cake cutting}), and on allocation of items that are not considered to be goods, but chores.
The fairness notions considered in this paper are {\em share based}. Other fairness notions, such as comparison based ones (envy freeness and its relaxations, such as EFX and EF1) or equity based ones (equitable allocations and its relaxation EQX) are not addressed in our work. Also, among share based notions, we only consider the MMS, APS and MES (which we regard as the most appealing ones), and do not consider others (such as WMMS or MXS). The interested reader can easily find more information on topics mentioned above (and others) by searching for them in the internet. Here we just comment that in the equal entitlement case, EF1 allocations exist for all classes of valuation functions~\cite{LiptonMMS04}, and likewise for MXS allocations~\cite{AR24}. Hence, also for subadditive valuations (and equal entitlements), EF1 allocations and MXS allocations always exist. However, these allocations might give only very poor guarantees with respect to the MMS. (Both EF1 and MXS are satisfied by envy free allocations, but for subadditive valuations, envy free allocations might give agents only $\frac{1}{\sqrt{m}}$-MMS.)

%We now turn to describe work more directly related to ours, for which we do provide references.

There has been relatively little work on ex-ante share based fairness. For additive valuations, the MES is the same as the proportional share, and it is trivial that each agent can get her proportional share ex-ante (give each agent $i$ all items with probability equal to her entitlement $b_i$). For XOS valuations and equal entitlements it is shown in~\cite{AkramiMSS23} (among other results) that one can give each agent ex-ante $\frac{1}{4}$-MMS (note that this refers to MMS, not MES), and no better than $\frac{3}{4}$-MMS. Our Theorem~\ref{thm:exante} provides much better (and tight) bounds.

There has been quite a lot of work on ex-post share based fairness. 
For additive valuations and equal entitlements, following a long line of work, it is known that $\frac{3}{4} + \frac{3}{3836}$-MMS allocations exist~\cite{akrami2023breaking}, and that there are instances in which no allocation gives more than $\frac{39}{40}$-MMS~\cite{FST21}. For arbitrary entitlements, $\frac{3}{5}$-APS allocations exist~\cite{BEF21APS}, shown using the bidding game. For submodular valuations and equal entitlements, $\frac{10}{27}$-MMS allocations exist~\cite{BUF23}, shown using the bidding game, and there are instances in which no allocation gives more than $\frac{2}{3}$-MMS~\cite{KKM23}. For arbitrary entitlements, $\frac{1}{3}$-APS allocations exist~\cite{BUF23}, shown using the bidding game.

Moving to XOS and subadditive valuations, the focus of our current paper, we present a more detailed account on related work. This related work concerns the equal entitlement case and the MMS. We are not aware of related work that presented any significant result for the unequal entitlements case and the APS.

MMS allocation for the equal entitlement case were first studied in~\cite{GhodsiHSSY22}. It was shown that there are instances in which no allocation gives more than $\frac{1}{2}$-MMS. This holds both for subadditive valuations and for XOS valuations. On the positive side, it was shown the $\frac{1}{5}$-MMS allocations exist for XOS valuations. {This implies that $\frac{1}{5\ln m}$-MMS allocations exist for subadditive valuations, because every subadditive valuation $v$ over a set $S$ can be lower bounded by an XOS valuation $v'$, with $v'(S) \ge \frac{1}{\ln m} v(S)$~\cite{BR11}.}

The approximation ratio for XOS valuations was somewhat improved in~\cite{SS24}, and then further improved to $\frac{3}{13}$-MMS~\cite{AkramiMSS23}. Our Theorem~\ref{thm:APSXOS4} improves over this, showing that $\frac{4}{17}$-APS allocations exist. 

For subadditive valuations and equal entitlements, the existence of {$\frac{3}{13\ln m}$-MMS} allocations follows from the results of~\cite{AkramiMSS23}.  Our Theorem~\ref{thm:biddingStrategy} improves over this, showing that $\frac{(1 - o(1))\log\log m}{\log m}$-APS allocations exist. In addition, following~\cite{SS24}, it is shown in~\cite{FH25} that $\frac{1}{14 \log n}$-MMS allocations exist. We note that even if $n$ is small, the number of items would have to be astronomical for this bound to improve over the bound in our Theorem~\ref{thm:biddingStrategy}. (And of course, the result of~\cite{FH25} does not apply to the arbitrary entitlement case.) 

Randomized allocation algorithms that simultaneously provide both good ex-ante guarantees and good ex-post guarantees are referred to as {\em best of both worlds} (BoBW) results. With equal entitlements, for additive valuations, one can get MES ex-ante simultaneously with $\frac{1}{2}$-MMS ex-post~\cite{BEF22BoBW}.   For XOS valuations, one can get $\frac{1}{4}$-MMS ex-ante (MMS, not MES) simultaneously with $\frac{1}{8}$-MMS ex-post~\cite{AkramiMSS23}.  In our work, we do not provide BoBW results.  

\section{Preliminaries}

In an allocation instance, $\items$ always denotes the set of items, $m$ denotes their number, and $n$ denotes the number of agents. The valuation function of agent $i$ is denoted by $v_i$, and her entitlement is denoted by $b_i$, where $b_i > 0$ and $\sum_i b_i = 1$. If $b_i = \frac{1}{n}$ for all agents, then the allocation instance is one of {\em equal entitlements}, and otherwise it is an allocation instance with {\em arbitrary entitlements}. Ex-post share based fairness notions (such as MMS and APS) refer to the value of the bundle received in the realized allocation, whereas ex-ante share based fairness notions (such as MES) refer to the expected value of the bundle received when the allocation procedure is randomized.

\subsection{The bidding game}
\label{sec:bidding}

The bidding game was introduced in~\cite{BEF21APS}, as an allocation procedure for agents with arbitrary entitlements. It comes in two versions, that we shall refer to as the {\em plain} version and the {\em extended} version.

Initially, each agent is given a budget equal to her entitlement, and all items are available for taking. Thereafter, the bidding game proceeds in rounds. In each round $r$, each agent places a bid of her choice, but not higher than her remaining budget. The agent placing the highest bid (breaking ties arbitrarily) wins the round. In the plain version of the bidding game, the agent pays her bid and selects an item of her choice, among the remaining items. In the extended version of the game, the winning agent may select more than one item, but needs to pay for each of the selected items. Her bid in that round serves as the price per item, and the number of items that she is allowed to select it upper bounded by the budget that she holds at that round -- the total payments of an agent are not allowed to exceed her budget.

When we refer to the bidding game in this work, we shall mean the extended bidding game, unless explicitly stated otherwise.

A {\em safe strategy} for the bidding game is a strategy that if used by an agent, provides some guarantee on the final value that the agent will receive, regardless of the strategies used by other agents. For example, a safe strategy that guarantees $\frac{3}{5}$-APS for agents with additive valuations was designed in~\cite{BEF21APS}. If all agents use the safe strategy, then all agents get the guarantee simultaneously, implying that an allocation with such a guarantee exists.

\subsection{The configuration LP}

A computational tool often used in the literature on maximum welfare allocations is the {\em configuration LP} (CLP, where LP stands for {\em linear program}). We first introduce the configuration IP (integer program). It has exponentially many variables, where variable $x_{i,S}$ (for agent $i$ and bundle $S \subset \items$) is intended to be a 0/1 indicator variable indicating whether the bundle received by agent $i$ is $S$. 

{\bf maximize} $\sum_{i,S} x_{i,S} \cdot v_{i}(S)$  (maximize welfare) subject to

\begin{itemize}
\item $\sum_S x_{i,S} = 1$ for every agent $i$ (every agent receives one bundle)
\item $\sum_{i,S \; \mid \; e \in S} x_{i,S} = 1$ for every item $e$ (every item is allocated once)
\item $x_{i,S} \in \{0,1\}$ for all $i,S$.
\end{itemize}

The configuration IP exactly characterizes the maximum welfare allocation.  Solving it is NP-hard. The configuration LP is a relaxation of the configuration IP, replacing the last constraint by  $x_{i,S} \ge 0$. As shown in~\cite{DNS10}, the configuration LP can be solved in polynomial time and a polynomial number of {\em demand queries}. A demand query to a valuation $v_i$ specifies a non-negative price $p_j$ for each item $j$, and gets in reply the bundle $S$ that maximizes $v_i(S) - \sum_{j\in S} p_j$ (breaking ties arbitrarily).

For the classes of valuations that most interest us in our work, XOS and subadditive, there are very useful rounding techniques that round any feasible solution of the configuration LP (the solution need not be optimal) to a feasible solution of the configuration IP, and hence, to a legal allocation~\cite{Feige09}. They lead to $1- \frac{1}{e} \simeq 0.632$ approximation to the maximum welfare in the case of XOS, and $\frac{1}{2}$ approximation for subadditive valuations. Moreover, they have additional properties that are useful in the context of fair allocations. We summarize the properties of these rounding techniques in the following theorems (both implicit in~\cite{Feige09}).

\begin{theorem}
    \label{thm:roundXOS}
    For XOS valuations, there is a randomized rounding procedure that for every feasible solution of the configuration LP produces a distribution over allocations with the following properties:
\begin{itemize}
\item
    For every agent $i$, the expected value (under $v_i$) of the bundle received by $i$ is at least a $(1 - \frac{1}{e})$ fraction of the agent's contribution $\sum_S x_{i,S} \cdot v_i(S)$ to the solution of the configuration LP.
    \item Moreover, for each bundle $S$ the rounding is such that $i$ tentatively gets the bundle $S$ with probability exactly $x_{i,S}$, and then, conditioned on the residual randomness of the rounding procedure, ends up with a bundle $S' \subseteq S$ of expected value at least $E[v_i(S')] \ge (1 - \frac{1}{e})v_i(S)$.
\end{itemize}
\end{theorem}

\begin{theorem}
    \label{thm:roundSubadditive}
    For subadditive valuations, there is a randomized rounding procedure that for every feasible solution of the configuration LP produces a distribution over allocations with the following properties:
\begin{itemize}
\item
    For every agent $i$, the expected value (under $v_i$) of the bundle received by $i$ is at least $\frac{1}{2}$ of the agent's contribution $\sum_S x_{i,S} v_i(S)$ to the solution of the configuration LP.
    \item Moreover, for each bundle $S$ the rounding is such that $i$ tentatively gets the bundle $S$ with probability exactly $x_{i,S}$, and then, conditioned on the residual randomness of the rounding procedure, ends up with a bundle $S' \subseteq S$ of expected value at least $E[v_i(S')] \ge \frac{1}{2}v_i(S)$. Moreover, $S'$ has probability at least $\frac{1}{2}$ of having value at least $\frac{1}{2}v_i(S)$.
\end{itemize}
\end{theorem}

\section{Ex-ante guarantees}

We prove here Theorem~\ref{thm:exante}, showing the existence of randomized allocations that are $\frac{1}{2}$-MES ex-ante for subadditive valuations, and $(1 - \frac{1}{e})$-MES ex-ante for XOS valuations.

%The following theorem is a relatively straightforward consequence of the results in~\cite{Feige09}.

\begin{proof}
The fractional MES partitions of the agents gives a feasible solution to the configuration LP, in which each agent contributes her MES value to the value of the solution. Applying the rounding of {Theorem}~\ref{thm:roundSubadditive}, this gives a randomized allocation in which the expected value received by every agent is at least half her MES in the subadditive case {(item 1(a) of Theorem~\ref{thm:exante}).} Applying the rounding of {Theorem}~\ref{thm:roundXOS}, this gives a randomized allocation in which the expected value received by every agent is at least a $1 - \frac{1}{e}$ fraction her MES in the XOS case {(item 2(a) of Theorem~\ref{thm:exante}).}

The proofs of 1(b) and 2(b) are variations on known examples showing integrality gaps for the configuration LP. Let the set of items be all vectors in $\{1, \ldots, n\}^n$. For every agent $i$, her MMS partition is into $n$ sets $B_1^i, \ldots, B_n^i$, where for every $j$, $B_j^i$ contains those vector whose $i$th coordinate has value $j$. Observe that in every allocation, at most one agent can get a bundle that contains a bundle from her MMS partition. 

In the subadditive case, for every agent $i$, we set the subadditive valuation $v_i$ as follows. For a nonempty $S\subseteq \items$, $v_i(S) = 2$ if $B_j^i \subseteq S$ for some $j$, and $v_i(S) = 1$ otherwise. Hence, the MMS value for every agent is~2, but in every allocation the total value received by all agents is at most $n+1$. Consequently, in every randomized allocation, some agent does not get expected value higher than $\frac{n+1}{2n}$ times her MMS.

In the XOS case, for every agent $i$ her XOS valuation function is {$v_i(S) = \max_j[\mid S \cap B_j^i \mid ]$.} Hence the MMS of every agent is $n^{n-1}$, whereas it can be seen that no allocation has welfare larger than $(1 - (\frac{n-1}{n})^n)\cdot n^n$. Hence in every randomized allocation, some agent does not get expected value higher than $(1 - (\frac{n-1}{n})^n)$ times her MMS. This ratio tends to $1 - \frac{1}{e}$ when $n$ grows.
\end{proof}

%Does~\cite{EzraFNTW19} showing integrality gap of~2 with two agents improve 1(b) to hold when $n=2$? No, because with two agents can give one at least the MMS, and the other at least $\frac{1}{2}v(\items)$ in expectation.

\section{Bidding with subadditive valuations}

%Throughout this section, a valuation function $v$ over a set $\items$ of items is a set function that is normalized ($v(\emptyset) = 0$), monotone non-decreasing, and not identically zero ($v(\items) > 0$).
In this section we prove Theorem~\ref{thm:biddingStrategy}, showing that with a subadditive valuation, agent $i$ has a strategy in the bidding game ensuring at least $(1 - o(1))\frac{\log\log m}{\log m}$-$APS(\items, v_i, b_i)$. The proof {is} divided into two parts.

\begin{proof}
{\bf (First part -- the one-shot bidding strategy.)}
Scale the subadditive valuation of agent $i$ by a multiplicative factor so that its APS has value~1. To achieve at least $\rho$-APS (for a value of $\rho$ to be determined later), we use the {\em one-shot} bidding strategy, which we now introduce. 

We say that a bundle $S$ is {\em acceptable} if $v_i(S) \ge \rho$. Agent $i$ is considered to be {\em active} as long as it holds her full budget $b_i$.
At a given round $r$ in which agent $i$ is active, let $\items^r$ denote the set of items still available in the beginning of the round, and let $S^r \subset \items^r$ denote the smallest acceptable bundle, breaking ties in favor of bundles of higher value. (If there is no acceptable bundle, the strategy is said to {\em fail}.) The agent bids $\frac{b_i}{|S^r|}$. If she wins the bid, she selects $S^r$ (by this achieving a value of $\rho$), pays her entire budget, and becomes inactive. If she does not win the bid, she remains active after round $r$.

If the strategy above does not fail, then necessarily there is a round in which the agent gets an acceptable bundle (thus attaining the desired value in ``one-shot"). It remains to determine the highest value of $\rho$ for which the above strategy cannot fail.

We now temporarily leave the proof Theorem~\ref{thm:biddingStrategy} (we shall return to it later), so as to develop some tools that will assist in its proof.
\end{proof}

%To implement the strategy of Theorem~\ref{thm:biddingStrategy}, it suffices that the agent knows her APS value, and that she can answer {\em constrained maximization queries} of the form: given a set $S$ and a parameter $t$, which set of $t$ items has the highest value. (Approximate answers to these queries give an approximation to the performance guarantee.) 

\begin{definition}
\label{def:ladder}
    Let $f$ be a valuation function over a set $\items$ of $m$ items. Given $0 \le \rho \le 1$,  let $G(f,\items,\rho)$ ($G$ for {\em greater}, or {\em good}) be the set of bundles in $\items$ of value at least $\rho \cdot f(\items)$ (that is, for every $S \in G(f,\items,\rho)$, $f(S) \ge \rho \cdot f(\items)$). Given a positive integer $k$, the {\em $k$-ladder} $L(f,\items,k)$ is a $k$ dimensional vector, where for every $1 \le j \le k$, $$L(f,\items,k)_j = \min_{S \in G(f,\items,j/k)} |S|.$$
\end{definition}

For example, if the 3-ladder of a valuation function $f$ is the vector $(2,4,7)$, this means that there is a set of two items of value at least $\frac{1}{3}f(\items)$, a set of four items of value at least $\frac{2}{3}f(\items)$, and a set of seven items of value $f(\items)$. Moreover, no single item has value at least $\frac{1}{3}f(\items)$, no set of three items has value at least $\frac{2}{3}f(\items)$, and no set of six items has value $f(\items)$,

Observe that for every valuation function over a set $\items$ of $m$ items and for every positive integer $k$, the entries of the $k$ ladder satisfy $L(f,\items,k)_1 \ge 1$, $L(f,\items,k)_k \le m$, and $L(f,\items,k)_{j+1} \ge L(f,\items,k)_{j}$ for all $1 \le j < k$.

The following lemmas present properties that hold for ladders of subadditive valuations.

\begin{lemma}
    \label{lem:ladderSuperadditive}
    For a valuation function $f$ over a set $\items$ of $m$ items, let $k$ be a positive integer, and consider the ladder $L(f,\items,k)$. Define the function $h_{f,k}$ over the domain $[k]$, where for every $j\in [k]$, $h_{f,k}(j) = L(f,\items,k)_{j}-1$. If $f$ is subadditive, then $h_{f,k}$ is superadditive (meaning that $h_{f,k}(i+j) \ge h_{f,k}(i)+h_{f,k}(j)$).
\end{lemma}

\begin{proof}
Suppose for the sake of contradiction that for some $i,j \in [k]$ (for which $i+j \in [k]$) it holds that $h_{f,k}(i+j) < h_{f,k}(i)+h_{f,k}(j)$. Then $L(f,\items,k)_{i+j} \le L(f,\items,k)_{i} - 1 + L(f,\items,k)_{j} - 1$. Let $S$ be set of size $L(f,\items,k)_{i+j}$ and value at least $\frac{i+j}{k} \cdot f(\items)$. Partition $S$ into a disjoint union of two sets, $S_1$ and $S_2$, where $|S_1| < L(f,\items,k)_{i}$ and $|S_2| < L(f,\items,k)_{j}$. Then $f(S_1) < \frac{i}{k} \cdot f(\items)$ and $f(S_2) < \frac{j}{k} \cdot f(\items)$, contradicting the subadditivity of $f$.
\end{proof}

\begin{lemma}
    \label{lem:ladderSubadditive}
    Let $f$ be a subadditive valuation function over a set $\items$ of $m$ items, let $k$ be a positive integer, and consider the ladder $L(f,\items,k)$. Then for every $1 \le j \le k-1$, for every set $S$ with $|S| < L(f,\items,k)_j$, it holds that $\items \setminus S$ contains a set $T$ of size $|T| \le L(f,\items,k)_{j+1}$ and value $v(T) > \frac{v(\items)}{k}$.
\end{lemma}

\begin{proof}
    Consider a set $U \in G(f,\items,(j+1)/k)$ of minimum size. Hence, $f(U) \ge \frac{j+1}{k} \cdot f(\items)$ and $|U| = L(f,\items,k)_{j+1}$. For $S$ as in the statement of the lemma, express $U$ as $U = (U \cap S) \cup T$, where $T = (U \setminus S)$. By subadditivity of $f$, $f(T) \ge f(U) - f(U \cap S)$. We can bound $f(U \cap S) \le f(S) < \frac{j}{k} \cdot f(\items)$, where the first inequality follows by monotonicity of $f$, and the last inequality is because $S$ is smaller than the smallest set in  $G(f,\items,j/k)$. Hence $f(T) > \frac{j+1}{k} \cdot f(\items) - \frac{j}{k} \cdot f(\items) = \frac{1}{k} \cdot f(\items)$, and $|T| \le |U| \le  L(f,\items,k)_{j+1}$.
\end{proof}

We now return to prove Theorem~\ref{thm:biddingStrategy}.

\begin{proof}
{\bf (Second part -- the analysis.)}
Recall the one shot bidding strategy from the first part 

\begin{claim}
\label{claim:tight}
    %Assuming Conjecture~\ref{conj:sum} below, 
    {If $k$ is an integer satisfying $m \le (k-1)^{k-1}$, the one-shot strategy succeeds for a value of $\rho$ satisfying $\rho = \frac{1}{k}$.}
    % $\rho = \frac{\log\log m}{(1 + o(1))\log m}$.
\end{claim}

\begin{proof}
We refer to the bundles of the APS fractional partition of agent $i$ as APS bundles. Consider an arbitrary APS bundle $B$ and a round $r$. Let $X^r$ denote the set of items consumed from $B$ (by other agents) up to the beginning of round $r$. If $v_i(B \setminus X^r) \ge \rho$, we say that bundle $B$ is still active in round $r$. In this case,  let $B^r \subset (B \setminus X^r)$ denote the smallest acceptable sub-bundle of $B \setminus X^r$.

{We set $\rho = \frac{1}{k}$ for an integer $k$ satisfying the conditions of the claim.} 
For an APS bundle $B$, consider the $k$-ladder $L(v_i,B,k)$ (recall Definition~\ref{def:ladder}). 
In a round $r$ in which $B$ is active, let  $j$ be the smallest index so that $|X^r| < L(v_i,B,k)_{j}$. If $j \le k-1$, then Lemma~\ref{lem:ladderSubadditive} implies that $|B^r| \le L(v_i,B,k)_{j+1}$. (We will not need to consider the case that $j = k$.) %\ufe{Moreover, if $\rho \le \frac{1}{2k}$, then subadditivity of $v_i$ implies that $|B^r| \le \lceil \frac{L(v_i,B,k)_{j+1}}{2} \rceil$.}

Suppose that during the bidding game, bundle $B$ became inactive. For $\ell \in \{1, \ldots, |B|\}$, let $p_{\ell}$ denote the price paid (by the agent who won the item) for the $\ell$th item consumed from $B$ (where $p_{\ell} = 0$ if fewer than $\ell$ items were consumed). This price is at least the bid of agent $i$ at the round in which the item was consumed. The bidding strategy implies that $p_1 \ge \frac{b_i}{L(v_i,B,k)_1}$. Moreover, for every $j \le k-1$, if $\ell \le L(v_i,B,k)_{j}$, then $p_{\ell} \ge \frac{b_i}{L(v_i,B,k)_{j+1}}$. Extend the $k$-ladder to coordinates~0 and $-1$, and set $L(v_i,B,k)_{0}= 1$ and $L(v_i,B,k)_{-1}= 0$. With this extended notation,~(\ref{eq:payments}) is a lower bound on how much the other agents paid for the items of $B$ until $B$ became inactive. 
\begin{equation}
\label{eq:payments}
   b_i \cdot \sum_{j=0}^{k-1} \frac{L(v_i,B,k)_{j} - L(v_i,B,k)_{j-1}}{L(v_i,B,k)_{j+1}} 
\end{equation}

We wish $k$ to be sufficiently large as a function of $m$ so that the value of the sum in~(\ref{eq:payments}) is at least~$1-b_i$.  To simplify the presentation, we shall derive bounds that hold for all values of $b_i$. Consequently, we shall want the sum to be at least~1, a more demanding requirement than at least $1 - b_i$. It will be more convenient for us to first fix $k$, and then see what is the maximum value $m_k$ such that for every $m \le m_k$ the sum in~(\ref{eq:payments}) is at least~$1$.

\begin{lemma}
\label{lem:sum}
    For every positive integer $k$, the sum in~(\ref{eq:payments}) is at least~1 for all $m \le m_k = (k-1)^{(k-1)}$. (Note that with these bounds, $k = (1 + o(1))\frac{\log m_k}{\log\log m_k}$.)
\end{lemma}

The proof of Lemma~\ref{lem:sum} is rather technical, and appears in section~\ref{sec:sumProof}.

With $k$ and $m$ as in Lemma~\ref{lem:sum}, other agents need to pay at least $b_i$ in order to make an APS bundle inactive. 
We claim that the total that other agents must pay in order to make all APS bundles inactive is at least~1. As the total budget of other agents is $1 - b_i$, this implies that the other agents cannot cause all APS bundles to become inactive. Consequently, there must be a round in which agent~$i$ wins, and gets a value of at least $\frac{1}{k}$.

It remains to prove our claim from the previous paragraph. Recall that the sum of weights $\lambda_j$ of the APS bundles is~1 (being at least~1 suffices for the conclusion), and that every item appears in APS bundles of total weight $b_i$ (total weight at most $b_i$ suffices for the conclusion).
Let $p_j$ be the price paid for item $e_j$. Consider a weighted sum the payments to all APS bundles, $\sum \lambda_j p(B_j)$. As $p(B_j) \ge b_i$ for each APS bundle and $\sum \lambda_j = 1$, we obtain that the value of this sum is at least $b_i$. As each item appears in bundles of total weight (at most) $b_i$, this sum is smaller than $b_i \sum_{j=1}^m p_j$. This implies  that $\sum_{j=1}^m p_j \ge 1$.

This completes the proof of Claim~\ref{claim:tight}.
\end{proof}

The above analysis of the one-shot strategy completes the proof of Theorem~\ref{thm:biddingStrategy}. 
\end{proof}

\begin{remark}
The one-shot strategy offers additional guarantees beyond that stated in Theorem~\ref{thm:biddingStrategy}.
For an agent with entitlement $b_i$, let $s$ be the smallest integer satisfying $b_i > \frac{1}{s+1}$, and let $e_s$ denote the $s$th most valuable item to agent $i$. Then the one-shot strategy ensures that the agent gets a value of at least $v_i(e_s)$. This is because if $\{e_s\}$ is an acceptable bundle, the agent bids $b_i$ as long as any of the top $s$ items remain, and other agents do not have sufficient budget to win all the top $s$ items.  
    %The proof of Theorem~\ref{thm:biddingStrategy} easily extends from APS to $\widehat{MMS}$, with only a somewhat worst constant in the $O$ notation. That is, for an agent with a subadditive valuation, there is a strategy in the multiple picks version of the bidding game that ensures the agent at least a $\frac{1}{k}$ fraction of her $\widehat{MMS}$, where $k = O(\frac{\log m}{\log\log m})$. The only difference in the proof is that we require the sum in~(\ref{eq:payments}) to be at least $\sum_{j \not= i} \hat{b}_j$ instead of $\sum_{j \not= i} {b}_j$. As we always have that $\sum_{j} \hat{b}_j < 1.7$, this changes only the constant hidden in the $O$ notation. 
\end{remark}

\section{Approximate APS allocations for XOS valuations}

In this section we show that there is no strategy for the bidding game that guarantees for agents with XOS valuations a constant {fraction} of their APS. Then we show other allocation algorithms that do offer such a guarantee.

\subsection{The bidding game for XOS valuations}

In this section we prove Theorem~\ref{thm:negative}, that in the case of equal entitlements and XOS valuations, no strategy in the  (extended) bidding game 
guarantees more than 
     $$(1 + \epsilon)\frac{\log\log m}{\log m} \cdot MMS(\items, v_i, \frac{1}{n}).$$ 
    %Moreover, in these instances the number of agents $n$  can be arbitrarily large, with the approximation ratio not being better better than $\sqrt{\frac{2}{\epsilon \cdot n}}$.

\begin{proof}
{Given $\epsilon > 0$,} we design an instance that involves two large integer parameters, $k$ and $q > k$, that will determine $n$ and $m$.

Agent $i$ has the following XOS valuation function $v_i$. The items are partitioned into $n$ equal size sets, $B_1, \ldots, B_n$. For every set $S \subset \items$, $v_i(S) = \max_{j} [v_i(S \cap B_j)]$, and $v_i$ is additive over the items of each set $B_j$ (this is an XOS valuation). 
Within each set $B_j$, there is one item of value~1, $q$ items of value $\frac{1}{q}$, $q^2$ items of value $\frac{1}{q^2}$, and so on, up to $q^{k-1}$ items of value $\frac{1}{q^{k=1}}$. Hence, $v_i(B_j) = k$ for all $j$, and $MMS(\items, v_i, \frac{1}{n}) = k$. We set $n = \frac{8kq}{\epsilon}$, and consequently $m \simeq \frac{8}{\epsilon} kq^k$. 

In the bidding game, it will be convenient for us to scale the budget of each agent so that every agent starts with a budget of $k$ instead of $\frac{1}{n}$. Clearly, executions in this scaled game are equivalent to executions in the unscaled bidding game (in which the initial budgets are $\frac{1}{n}$), by scaling all bids by the same factor. 

When we refer to the {\em value} of an item, we mean its value under $v_i$. 

We partition the agents other than $i$ into two sets, $P_1$ that contains $\frac{n}{2}$ agents, and $P_2$ that contains $\frac{4kq}{\epsilon} - 1$ agents. 

%We present the strategy of the other agents, that prevents agent $i$ from obtaining a bundle of value more than $O(1)$.  in round $r$. The strategy of the other agents has two modes of operation, one executed by agents of $P_1$, and the other by agents of $P_2$. The choice of which of these two modes to use depends on the bid of agent $i$. 

In a round $r$, let $t_r$ denote the highest value  of an item that is still available. The bid of an agent from $P_1$ in round $r$ is $\frac{t_r}{2}$.
If the bid of agent~$i$ in round $r$ is less than $\frac{t_r}{2}$, then an agent from $P_1$ wins the round, and takes an item of value $t_r$. As there are $n/2$ agents in $P_1$, they can afford to buy all items at half their value. Consequently, for each item that agent $i$ wins, she needs to pay at least half its value.

The agents in $P_2$ bid in round $r$ as follows. An MMS bundle is referred to as {\em dangerous} if agent $i$ already has items of total value at least $\frac{\epsilon}{2}$ from this bundle. If there are no dangerous bundles, then the agents in $P_2$ bid~0. Else, let $s_r$ denote the highest value of an item among those remaining in dangerous bundles. An agent in $P_2$ bids $q \cdot s_r$. If she wins the bid, she takes an item of value $s_r$ from a dangerous bundle.

Observe that to take all remaining items in a dangerous bundle, agents in $P_2$ need to pay $(k-\frac{\epsilon}{2})q$. Agent $i$ has a budget of $k$, so she can create at most {$\frac{4k}{\epsilon}$} dangerous  bundles {(because to win an item, agent $i$ needs to bid at least half its value).} Hence, a budget of $\frac{4k}{\epsilon} \cdot (k-\frac{\epsilon}{2}) q$ suffices in order to carry out the strategy of agents in $P_2$. As their total budget is $(\frac{4kq}{\epsilon} - 1)k$, they can carry out their strategy without running out of budget.

We now upper bound the value that agent $i$ can reach. Let $B$ be an MMS bundle that became dangerous. In the round $r$ that $B$ became dangerous, agent $i$ could take at most $\frac{1}{t_r}$ items of value $t_r$ in $B$ (there are no more such items), and at most $\frac{2k}{t_r}$ items of value $\frac{t_r}{q}$ from $B$ (as her bid is at least $\frac{t_r}{2}$, and her total budget is $k$). In future rounds, the bids of $P_2$ limit the additional value that agent $i$ can take from $B$ to $\frac{k}{q}$, as agent $i$ needs to pay for these items $q$ times their value. Before round $r$, agent $i$ had a value of at most $\frac{\epsilon}{2}$. Hence altogether, agent $i$ gets items of total value at most $1 + \frac{3k}{q} + \frac{\epsilon}{2}$ from each individual MMS bundle.  As $v_i$ takes the maximum over the contributions of the MMS bundles, the total value that agent $i$ gets is also $1 + \frac{3k}{q} + \frac{\epsilon}{2}$. 

For a desired small value $\epsilon > 0$, we set $q =\frac{12k}{\epsilon}$, and then agent $i$ is limited to a value of $1 + \frac{3\epsilon}{4}$, which is $\frac{1+3\epsilon/4}{k}  MMS(\items, v_i, \frac{1}{n})$. By making $k$ sufficiently large, we can enforce $k > \frac{4 + 3\epsilon}{4 + 4\epsilon}\frac{\log m}{\log\log m}$, and then $i$ gets value of at most $(1 +{\epsilon})\frac{\log m}{\log\log m} MMS(\items, v_i, \frac{1}{n})$.

Recall also that $n = \frac{8kq}{\epsilon} = \frac{96k^2}{\epsilon^2}$, implying that $k = \Omega(\epsilon \sqrt{n})$. Fixing $\epsilon$ and letting $n$ grow, this implies that agent $i$ gets at most $O(\frac{1}{\sqrt{n}}) MMS(\items, v_i, \frac{1}{n})$  
\end{proof}

\subsection{Some general tools}

Theorem~\ref{thm:APSXOS} and Corollary~\ref{cor:APSXOS} are both extensions of an approach developed in~\cite{GhodsiHSSY22}. Our extension is in considering fractional partitions of the items (as allowed in the definition of the APS), whereas previous uses of his approach used integral partitions (as required in the definition of the MMS). 

\begin{theorem}
\label{thm:APSXOS}
    Consider an allocation instance for a set $\items$ of indivisible items with $n$ agents, where each agent $i$ has an XOS valuation $v_i$ and entitlement $b_i$,  with $b = \sum_{i=1}^n b_i$  (we do not require here that $b = 1$). Then there is an allocation in which each agent $i$ gets a bundle of value at least $(1 - b + b_i)APS(\items,v_i,b_i)$. 
\end{theorem}

\begin{proof}
Without loss of generality, we may assume that the APS of every agent is positive. For each agent $i$, define a valuation $\hat{v}_i$ that satisfies for every $S \subseteq \items$:
$$\hat{v}_i(S) = \min[b_i, \frac{b_i}{APS(\items,v_i,b_i)}v_i(S)].$$
Observe that the fractional APS partition for $\hat{v}_i$ is identical to that for $v_i$, and the APS value becomes $b_i$. The theorem will be proved if we show that each agent $i$ gets a bundle of $\hat{v}_i$ value at least $(1 - b + b_i)b_i$.

Consider an allocation $A = A_1, \ldots, A_n$ maximizing the welfare $\sum_i \hat{v}_i(A_i)$. We claim that each bundle $A_i$ satisfies $\hat{v}_i(A_i) \ge (1 - b + b_i)b_i$. For the sake of contradiction, suppose that this fails to hold for agent $i$. Then consider the following distribution over allocations. Sample a new bundle $B_i$ at random from the distribution implied by the fractional APS partition of $\hat{v}_i$. This produces an allocation $B$ in which agent $i$ receives $B_i$ instead of $A_i$, and every other agent $j$ receives the bundle $B_j = A_j \setminus B_i$. Then $$\hat{v}_i(B_i) - \hat{v}_i(A_i) > b_i - (1 - b + b_i)b_i = (b - b_i)b_i$$ Thus the welfare of agent $i$ increased by more than $(b - b_i)b_i$.

By the definition of the fractional APS partition, each item is in $B_i$ with probability $b_i$ (though items may be correlated). As the valuations are XOS, this implies that in expectation (over choice of $B_i$), for every agent $j \not=i$ it holds that $E[\hat{v}_j(B_j)] \ge (1 - b_i)\hat{v}_j(A_j)$. Hence in expectation, all other agents combined lost a value of at most 
$$\sum_{j \not= i} \left(\hat{v}_j(A_j) - E[\hat{v}_j(B_j)]\right) \le \sum_{j \not= i} b_i \hat{v}_j(A_j) \le \sum_{j \not= i} b_i b_j = b_i(b - b_i)$$

As agent $i$ gains more than the expected loss of all other agents, there must be a choice of $B_i$ that leads to an allocation $B$ with higher welfare than that of $A$, which contradicts the requirement that $A$ maximizes welfare.    
\end{proof}

%\begin{corollary}
%\label{cor:napprox}
%    Consider an allocation instance for a set $\items$ of indivisible items with $n$ agents, where each agent $i$ has an XOS valuation $v_i$ and entitlement $b_i$,  with $\sum_{i=1}^n b_i = 1$. Then there is an allocation in which each agent $i$ gets a bundle of value at least $b_i \cdot APS(\items,v_i,b_i)$. In particular, if agents have equal entitlements,  
%\end{corollary}

In the case of interest, when $\sum_i b_i = 1$, there is at least one agent $i$ with $b_i \le \frac{1}{n}$, and Theorem~\ref{thm:APSXOS} does not offer this agent a guarantee better than $\frac{1}{n} \cdot APS(\items,v_i,b_i)$. To obtain approximations of the APS that are independent of $n$, we shall use the following corollary of Theorem~\ref{thm:APSXOS}. 

\begin{corollary}
\label{cor:APSXOS}
    Consider an allocation instance for a set $\items$ of indivisible items with $n$ agents, where each agent $i$ has an XOS valuation $v_i$ and entitlement $b_i$,  with $\sum_{i=1}^n b_i = 1$. Then the following hold. 
    \begin{enumerate}
        \item There is an allocation in which each agent $i$ gets a bundle of value at least $\frac{1+b_i}{2} APS(\items,v_i,\frac{b_i}{2})$.
        \item For each $i$ let $0 < \alpha_i \le 1$ be such that no single item $e$ has value $v_i(e) > \alpha_i \cdot APS(\items,v_i,b_i)$. Then there is an allocation in which each agent $i$ gets a bundle of value at least $\frac{1- \alpha_i + b_i}{4}{APS(\items,v_i,b_i})$.
    \end{enumerate} 
    \end{corollary}

    \begin{proof}
        Item~1 of the corollary follows by noting that for entitlements as in the corollary $\sum_i \frac{b_i}{2} = \frac{1}{2}$. Hence, we are in the setting of Theorem~\ref{thm:APSXOS} with $b = \frac{1}{2}$.

        For item~2, consider an arbitrary bundle $B$ in the fractional $APS(\items,v_i,b_i)$ partition of agent $i$ (hence $v_i(B) \ge APS(\items,v_i,b_i)$), and let $w$ denote its weight. Partition $B$ into two bundles, $B_1$ and $B_2$, by inserting items of $B$ into $B_1$ until its value reaches or exceeds {$\frac{1- \alpha_i}{2}{APS(\items,v_i,b_i})$}. At that point, 
        $$\frac{1- \alpha_i}{2}APS(\items,v_i,b_i) \le v_i(B_1) < \frac{1+ \alpha_i}{2}APS(\items,v_i,b_i)$$
        \noindent because no item has value larger than $\alpha_i \cdot APS(\items,v_i,b_i)$.
        Subadditivity of $v_i$ implies that $v_i(B_2) > \frac{1- \alpha_i}{2}{APS(\items,v_i,b_i}$. Give each of $B_1$ and $B_2$ a weight of $\frac{w}{2}$. Doing this process for each bundle in the fractional $APS(\items,v_i,b_i)$ partition gives a fractional $APS(\items,v_i,\frac{b_i}{2})$ partition in which each bundle has value at least $\frac{1- \alpha_i}{2}APS(\items,v_i,b_i)$.  Combining this with item~1 of the corollary, we get item~2.
    \end{proof}

If $\alpha_i \le b_i$ in item~2 of Corollary~\ref{cor:APSXOS}, then the allocation gives the agent a value of at least $\frac{1}{4} \cdot APS(\items,v_i,b_i)$. However, it might be that $\alpha_i$ is as large as~1 (recall that we may assume without loss of generality that $\alpha_i \le 1$), or nearly~1, in which cases Corollary~\ref{cor:APSXOS} does not seem to offer significant advantages over direct use of Theorem~\ref{thm:APSXOS}.

To handle the case that there are items of value nearly as large as the APS, we wish to treat items of large value separately. However, we do not know how to do this for input instances with arbitrary entitlements. For this reason, we shall present two allocation algorithms, one for the arbitrary entitlements case, and one for equal entitlements.

\subsection{Arbitrary entitlements}

Here we prove Theorem~\ref{thm:APSXOS6}, showing that $\frac{1}{6}$-APS allocations exist when agents have XOS valuations and arbitrary entitlements. It may be helpful if readers familiarize themselves with the proof of Theorem~\ref{thm:APSXOS} before reading the following proof.

\begin{proof}
Without loss of generality, we may assume that the APS of every agent $i$ is positive and equal to $b_i$ (this last property can be achieved by scaling the valuation). An allocation is {\em acceptable} if every agent $i$ gets a bundle of $v_i$ value at least $\frac{b_i}{6}$. For each agent $i$, define a valuation $\hat{v}_i = \min[v_i, \frac{b_i}{3}]$. Note that in an acceptable allocation, it is also true that every agent $i$ gets a bundle of $\hat{v}_i$ value at least $\frac{b_i}{6}$.

%that satisfies for every $S \subseteq \items$:
%$$\hat{v}_i(S) = \min[\frac{b_i}{3}, v_i(S)].$$
%The theorem will be proved if we show that each agent $i$ gets a bundle of $\hat{v}_i$ value at least $\frac{b_i}{6}$. We refer to such an allocation as an {\em acceptable allocation}.  

Assume for the sake of contradiction that there is no acceptable allocation. We claim that %this contradicts the fact that there are finitely many possible allocations. Specifically, we show that 
this implies that for every allocation $A = A_1, \ldots, A_n$ there is an allocation $A' = A'_1, \ldots, A'_n$ with strictly higher $\hat{v}$ welfare, namely, $\sum_i \hat{v}_i(A'_i) > \sum_i \hat{v}_i(A_i)$. Consequently, the assumption leads to the conclusion that there are infinitely many possible allocations, contradicting the fact that the number of possible allocations is finite. Note that our proof does not imply that the allocation that maximizes $\hat{v}$ welfare is acceptable (unlike the case in the proof of Theorem~\ref{thm:APSXOS}).

We now prove our claim that if there is no acceptable allocation, then for every allocation $A$ there is an allocation $A'$ with higher $\hat{v}$ welfare. Consider an allocation $A = A_1, \ldots, A_n$. By our assumption, there is an agent $i$ for which $\hat{v_i}(A_i) < \frac{b_i}{6}$. Consider her fractional $APS(\items,v_i,b_i)$ partition. We replace this fractional partition by a new fractional partition as follows.

Let $B$ be a bundle in the support of the fractional partition, and let $\lambda$ be its weight.  If $B$ contains a {\em large} item $e$, one of value $v_i(e) \ge \frac{b_i}{3}$, then replace $B$ by the bundle $B' = \{e\}$, and let $B'$ maintain the weight $\lambda$. (If there is more than one such possible $e$, then choose one of them arbitrarily.) 
%We remark that we can choose any $t$ in the range $\frac{b_i}{3} \le t \le \frac{2b_i}{3}$ in the definition of large value $v_i(e) \ge t$, and our proof would still go through.) 
We refer to $B'$ as a {\em single-item} bundle.
If there is no large $e$, then partition $B$ into two bundles, $B'_1$ and $B'_2$, each of value at least $\frac{b_i}{3}$. (This can be done similarly to the proof of item~2 in Corollary~\ref{cor:APSXOS}.) Give each of these two bundles weight $\frac{\lambda}{2}$. We refer to these bundles as {\em multiple-items} bundles.

Let $\alpha_i$ denote the total weight of single-item bundles in this new fractional partition. Hence $1 - \alpha_i$ is the total weight of multiple-items bundles. (Note that these weights depend only on agent $i$, not on the allocation $A$.)

In the allocation $A$, let $W_{-i} = \sum_{j \not= i} \hat{v}_j(A_j)$ denote the total $\hat{v}$ welfare of agents other than $i$, and note that $W_{-i} \le \sum_{j \not= i} \frac{b_j}{3} = \frac{1 - b_i}{3}$. Let $\alpha$ denote the fraction of this welfare that is contributed by 
those items that make up the single-item bundles of agent $i$. (To make this contribution well defined, we do as follows. For each agent $j$, her XOS valuation $\hat{v}_j$ over $A_j$ is replaced by an additive valuation  $v'_j$ so that $v'_j(A_j) = \hat{v}_j(A_j)$ and $v'_j(S) \le \hat{v}_j(S)$ for every $S \subset A_j$. The contribution to the welfare of an item $e \in A_j$ is $v'_j(e)$.)

Now we consider two cases. 

\begin{enumerate}
    \item If $\alpha_i \ge \alpha$ agent $i$ replaces her bundle $A_i$ by a single-item bundle $B'$, where item $e \in B'$ is the item of smallest contribution to $W_{-i}$ among the items that make up single-item bundles. As this contribution is no larger than the average contribution if we select $e$ at random by the distribution induced by the new fractional partition on the single-item bundles, the loss in $\hat{v}$ welfare to the other agents is at most $\alpha W_{-i} \cdot \frac{b_i}{\alpha_i} \le b_i W_{-i} \le  \frac{b_i (1 - b_i)}{3}$. 
    \item If $1 - \alpha_i > 1 - \alpha$ agent $i$ replaces her bundle $A_i$ by a multiple-items bundle $B'$, where $B'$ has the set of items of smallest contribution to $W_{-i}$ among those sets that make up multiple-item bundles. %As this contribution is no larger than the average contribution when select $B'$ at random by the distribution induced by the new fraction partition on the multiple-items bundles,
    The loss in $\hat{v}$ welfare to the other agents is at most $(1 - \alpha) W_{-i} \cdot \frac{b_i}{2(1 - \alpha_i)} \le  \frac{b_i (1 - b_i)}{6}$.
    %where the last inequality follows because $W_{-i} = \sum_{j \not= i} \hat{v}_j(A_j) \le \sum_{j \not= i} \frac{b_j}{3}$.  
\end{enumerate}

The new allocation obtained after applying the appropriate case above will be called $A^1$. Allocation $A^1$ will be referred to as an allocation of type~1 if it was reached by applying case~1 above, and of type~2 if it was reached by applying case~2 above.

Let us compare the $\hat{v}$ welfare of $A^1$ with that of $A$. The gain in $\hat{v}_i$ welfare for agent $i$ is $\hat{v}_i(B') - \hat{v_i}(A_i) = \frac{b_i}{3} - \hat{v_i}(A_i)$. For the other agents, the loss in welfare is at most $\frac{b_i (1 - b_i)}{3}$ if $A^1$ of type~1, and at most $\frac{b_i (1 - b_i)}{6}$ if $A^1$ of type~2. As $\hat{v_i}(A_i) < \frac{b_i}{6}$, if $A^1$ is of type~2  then the agents gain welfare, and we can take $A' = A^1$. However, if $A^1$ is of type~1, the agents might lose welfare. 

To handle the above possible loss, we create a sequence $A^1, A^2, \ldots$ of allocations. Each new allocation $A^{j+1}$ is derived from the previous allocation $A^j$ by the same procedure as above: identifying an agent that does not receive an acceptable bundle in $A^j$ (such an agent must exist, by our assumption that there are no acceptable allocations), and applying either case~1 or~2 as above. Hence, the sequence can be of any length that we desire. We now track the change in welfare over the whole sequence, and show that at some point the welfare must be larger than that of $A$.

Partition the sequence into $n$ subsequences, one for each agent. Subsequence $A^{i_1}, A^{i_2}, \ldots$ for agent $i$ contains all those indices $j$ for which agent $i$ was the one to give up her bundle $A^{j-1}_i$ in allocation $A^{j-1}$, and received an alternative bundle $B'$ instead (as in one of the two cases described above), creating $A^j$. Let us track the changes in welfare that result from the subsequence of $i$. We use the observation that if $A^{i_j}$ is an allocation of type~1, then in allocation $A^{i_{j+1}-1}$ agent $i$ does not have any item. This is because to reach a value smaller than $\frac{b_i}{6}$, agent $i$ must have lost the only item that she held in $A^{i_j}$. 

Suppose that subsequence $i$ has $t_1$ indices with allocations of type~1, and $t_2$ indices with allocations of type~2.

Then other agents lost at most $t_1\cdot \frac{b_i (1 - b_i)}{3} + t_2\cdot \frac{b_i (1 - b_i)}{6}$. Agent~1 gains at least $(t_1 + t_2) \cdot \frac{b_i}{3} - (t_2 + 1)\cdot \frac{b_i}{6}$. (We have a term of $(t_2 + 1)\cdot \frac{b_i}{6}$ rather than $t_2 \cdot \frac{b_i}{6}$ because the last allocation in the subsequence might be of type~1.) Altogether, over the whole subsequence the gain in welfare is at least $(b_i)^2(\frac{t_1}{3} + \frac{t_2}{6}) - \frac{b_i}{6}$. Hence, there cannot be a loss larger than $\frac{b_i}{6}$, and the gain tends to infinity as the length of the subsequence grows.

Summing up the contributions of the subsequences of all agents we see that if the sequence $A^1, A^2, \ldots$ contains more than $(\frac{1}{b_n})^2$ terms (where $b_n$ is the smallest entitlement), then we necessary reach an allocation with higher welfare than that of $A$. 
\end{proof}

\subsection{Equal entitlements}

Here we prove Theorem~\ref{thm:APSXOS6}, showing that $\frac{4}{17}$-APS allocations exist when agents have XOS valuations and arbitrary entitlements.

Let us recap the allocation instances that our next algorithm will handle. There are $n$ agents of equal entitlements. The APS of every agent is positive, and valuation functions are scaled so that the APS of every agent is~1. We wish to give each agent an {\em acceptable} bundle, where being acceptable means that its value for the agent is at least $\rho$ (the approximation ratio). Our algorithm is as follows.

{\bf Allocation algorithm:}

\begin{enumerate}
    \item Let $S$ be the smallest set of items (breaking ties arbitrarily) that is acceptable to some agent. 
    \begin{enumerate}
        \item If $|S| \le 4$, give $S$ to an agent $i$ for which $S$ is acceptable (among such agents, the choice can be arbitrary). Remove $S$ from the set of items, remove $i$ from the set of agents, and repeat step~1.
        \item If $|S| \ge 5$, move to step~2. 
    \end{enumerate} 
    \item Let $\items_5$ and $\agents_5$ be the sets of items and agents remaining at the beginning of this step, and let $n_5$ denote the number of these agents. As in item~1 of Corollary~\ref{cor:APSXOS}, allocate $\items_5$ to $\agents_5$ using an allocation that gives each agent $i \in \agents_5$ a bundle of value at least $\frac{1}{2} \cdot APS(\items_5, v_i, \frac{1}{2n_5})$. (The ratio $\frac{1}{2}$ can be replaced by $\frac{n_5 + 1}{2n_5}$, but we omit this minor improvement, for simplicity.)
\end{enumerate}

Clearly, every agent in $\agents \setminus \agents_5$ gets an acceptable bundle. It remains to show that also
every agent in $\agents_5$ gets an acceptable value. This requires a careful choice of the approximation ratio $\rho$.

\begin{lemma}
    \label{lem:417}
    If $\rho = \frac{4}{17}$ and bundles are acceptable if their value is at least $\rho$, then the allocation algorithm gives every agent in $\agents_5$ an acceptable bundle. In other words, setting $\rho = \frac{4}{17}$, the allocation algorithm gives every agent $i$ a bundle of value at least $\frac{4}{17} \cdot APS(\items, v_i, \frac{1}{n})$. 
\end{lemma}

Before proving Lemma~\ref{lem:417}, we compare our allocation algorithm with previous allocation algorithms for XOS valuations. Our algorithm is patterned after these earlier algorithms, with some differences. Our step~2 uses Corollary~\ref{cor:APSXOS}, which follows from Theorem~\ref{thm:APSXOS}. As noted earlier, this theorem (and corollary) is an extension of an approach developed in~\cite{GhodsiHSSY22}. The extension handles fractional partitions and not just (integral) partitions. For this reason, previous allocation algorithms approximate only the MMS, whereas our allocation algorithm approximates the APS (which is at least as large as the MMS, and may be larger). Another difference is in step~1 of the algorithm. We allow sets of size up to~4. In~\cite{GhodsiHSSY22}, the allowed size was only~1, leading to an approximation ratio of $\frac{1}{1 \cdot 4 + 1} = \frac{1}{5}$. In~\cite{AkramiMSS23} the allowed size went up to~3,  leading to an approximation ratio of $\frac{3}{3 \cdot 4 + 1} = \frac{3}{13}$. We remark that with the same type of proofs, with an allowed size of~2, the approximation ratio would be $\frac{2}{2 \cdot 4 + 1} = \frac{2}{9}$. Our switch from integral partitions to fractional partitions gives us more flexibility in our proofs, and allows us to go up to sets of size~4, for which we show an approximation ratio of $\frac{4}{4 \cdot 4 + 1} = \frac{4}{17}$.

We break to proof of Lemma~\ref{lem:417} to intermediate lemmas.

\begin{lemma}
\label{lem:fractionalPartition}
    Let $v$ be an additive valuation over bundle $B$ with $v(B) = 1$. Suppose that for every set $S \subset B$ with $|S| = 2$ it holds that $v(S) < \frac{1}{4}$. Then for every item $e\in B$, there are three sets $B_1, B_2, B_3 \subset (B \setminus \{e\})$ such that $v(B_j) \ge \frac{1}{2}$ for every $j \in \{1,2,3\}$, and every item $e' \in (B \setminus \{e\})$ is in at most two of the three sets.
\end{lemma}

\begin{proof}
Let $e$ be the item removed from $B$. Note that every item $e' \not= e$ satisfies $v(e') < \frac{1}{4} - v(e)$. 

    Partition $B \setminus \{e\}$ into three disjoint sets $S_1, S_2, S_3$ satisfying $v(S_1) \ge v(S_2) \ge v(S_3)$, and minimizing $v(S_1) - v(S_3)$. We claim that $v(S_1) \le \frac{1}{2} - v(e)$ (implying that $v(S_2 \cup S_3) \ge \frac{1}{2}$, because $v(S_2 \cup S_3) + v(S_1) + v(e) = 1$). The claim follows because otherwise $v(S_3) < \frac{1}{4}$ and for every $e' \in S_1$ it holds that $v(S_1 \setminus \{e'\}) > \frac{1}{2} - v(\{e\} \cup \{e'\}) > \frac{1}{4}$. Moving any $e' \in S_1$ with nonzero value to $S_3$ would demonstrate that the original choice of $S_1,S_2,S_3$ was not the one that minimizes $v(S_1) - v(S_3)$.

    The sets $B_1 = S_1 \cup S_2$, $B_2 = S_2 \cup S_3$, and $B_3 = S_3 \cup S_1$ satisfy the lemma.
\end{proof}

\begin{lemma}
\label{lem:817}
    Let $\items_5$ and $\agents_5$ denote the sets of items and agents that participate in step~5 of the allocation algorithm, and let $n_5 = |\agents_5|$. Then for every agent $i \in \agents_5$ it holds that $APS(\items_5, v_i, \frac{1}{2n_5}) \ge \frac{8}{17}$.
\end{lemma}

\begin{proof}
    %For $j \in \{1,2,3,4\}$ let $n_j$ denote the number of agents that took $j$ items. Let $\agents_5$ denote the set of remaining agents, and let $n_5 = n - n_1 - n_2 - n_3 -n_4$ denote their number. We may assume that $n_1 = 0$, because removing any agent $j$ and any single item $e$, the APS value of all other agents does not decrease (each agent can remove from her fractional APS partition the bundles containing $e$, and scale the \ufe{weights} of the remaining bundles by a factor of $\frac{1}{1 - b_j}$). 
Consider any agent $i \in \agents_5$ and any bundle $B$ in her fractional APS partition, and let $w$ denote its weight. Recall that we assume that $v_i(B) = 1$. %Observe that $v_i(e) < \frac{4}{17}$ for every $e \in B$, because $n_1 = 0$. 
Let $m_1, m_2, m_3, m_4$ denote the number of items taken from $B$ in executions of step~1 of the allocation algorithm, in which the respective allocated set $S$ was of size~1,~2,~3 or~4, respectively. Let $B'$ denote the set of items that remain in $B$ after step~1. The case analysis below determines the weight attributed to $B'$. (For each bundle $B$, the first case that applies is used. Cases~3 and~4 are somewhat more inclusive than what is actually needed for our proof. In them, it would suffice to replace the weak upper bounds on $m_3 + m_4$ by strict upper bounds, as the cases with equality in the upper bounds are handled by case~1.)

%WARNING: the text after the enumeration refers to these numbers. Change with care

\begin{enumerate}

\item $\frac{1}{m_1} + \frac{1}{m_2} + \frac{1}{m_3} + \frac{1}{m_4} \ge 1$. In this case, we set the weight of $B'$ to~0. 

%\item $m_1 \ge 1$. In this case, we set the weight of $B'$ to~0. 
    
    \item $m_1 + m_2 + m_3 + m_4 = 0$ (implying $B' = B$). In this case, the set of four top items in $B$ has value less than $\frac{4}{17}$. This implies that each of the remaining items has value at most $\frac{1}{17}$. Consequently, a proof similar to that of item~1 in Corollary~\ref{cor:APSXOS} shows that $B$ can be partitioned into two disjoint sets, $B_1$ and $B_2$, each of value at least $\frac{8}{17}$. Set the weight of each of $B_1$ and $B_2$ to $\frac{w}{2} \cdot \frac{n}{n_5}$. Hence, the parts of $B'$ contribute weight $w \cdot\frac{n}{n_5}$.

%    \item $m_2 \ge 2$. In this case, we set the weight of $B'$ to~0.
    
\item $m_2 = 1$ and $m_3 + m_4 \le 2$. Necessarily, $v_i(B') > \frac{1}{2}$. (The item taken at step~2 had value smaller than $\frac{4}{17}$, and the additional two items taken also had total value smaller than $\frac{4}{17}$, as otherwise agent $i$ would have taken two items.) Set the weight of $B'$ to $\frac{w}{2} \cdot \frac{n}{n_5}$. %Hence, the bundle contributes weight $\frac{w}{2} \cdot \frac{n}{n_5}$.

%\item $m_2 = 1$ and $m_3 + m_4 \ge 3$. In this case, we set the weight of $B'$ to~0.
%    \item $m_2 = 0$ and $m_3 + m_4 \ge 5$. In this case, we set the weight of $B'$ to~0.
\item $m_2 = 0$ and $2 \le m_3 + m_4 \le 4$. 
Necessarily, $v_i(B') > \frac{1}{2}$. (No two items in $B$ had total value at least $\frac{4}{17}$, as otherwise agent $i$ would have taken two items.) Set the weight of $B'$ to $\frac{w}{2} \cdot \frac{n}{n_5}$. %Hence, the bundle contributes weight $\frac{w}{2} \cdot \frac{n}{n_5}$.
     \item $m_2 = 0$ and $m_3 + m_4 = 1$. No two items in $B$ had total value at least $\frac{4}{17}$, as otherwise agent $i$ would have taken two items. Let $e$ be the item removed from $B$.  
     Then $B' = B \setminus \{e\}$ contains three bundles, $B_1, B_2, B_3$, with properties as in Lemma~\ref{lem:fractionalPartition}. Set the weight of each of these three bundles to $\frac{w}{4} \cdot \frac{n}{n_5}$. Hence, the parts of $B'$ contribute weight $\frac{3w}{4} \cdot \frac{n}{n_5}$.
\end{enumerate}

Observe that with the above weights, each item is in bundles of weight at most $b_i \cdot \frac{1}{2} \cdot \frac{n}{n_5} = \frac{1}{2n} \cdot \frac{n}{n_5} = \frac{1}{2n_5}$, and each generated bundle ($B'$, or its parts in cases~2 and~5) has value at least $\frac{8}{17}$ (the worst case is case~2). 

It remains to show that the sum of weights is at least~1. For this purpose, we refer to the weights in the case analysis above as the {\em desired} weights. In addition, we shall compute {\em pessimistic} weights that will assist in the analysis.  
%We show that the pessimistic weights sum up to at least~1, and that for every bundle, its desired weight is at least as high as its pessimistic weight. 

We associate with each bundle $B$ an initial pessimistic weight equal to its weight $w$ in the original fractional APS partition. Thus initially, the pessimistic weights sum up to at least~1. With each agent that takes items, we associate a reduction of at most~$\frac{1}{n}$ in the sum of pessimistic weights. This is done as follows. If the agent took $j$ items (for $j \in \{1,2,3,4\}$), then each such item reduces $\frac{1}{j}\cdot w$ from the pessimistic weight $w$ of the bundle to which it belongs. As each item belongs to bundles of total weight at most $\frac{1}{n}$, the total reduction in weights is at most $\frac{1}{n}$. Hence, when only agents of $\agents_5$ remain, the sum of pessimistic weights is at least $\frac{n_5}{n}$. Then, we scale the pessimistic weights by a multiplicative factor of $\frac{n}{n_5}$, reaching a sum of  weights that is at least~1. Going over the five cases above, we see that for every bundle, the desired weight is at least as high as the scaled pessimistic weight. Consequently, the sum of desired weights is at least~1.
\end{proof}

The proof of Lemma~\ref{lem:417} follows easily by plugging in the lower bound on $APS(\items,v_i,\frac{b_i}{2})$ given in Lemma~\ref{lem:817} into the approximation ratio given in item~1 of Corollary~\ref{cor:APSXOS}.

\subsection*{Acknowledgements}

This research was supported in part by the Israel Science Foundation (grant No. 1122/22).

\bibliographystyle{alpha}

\newcommand{\etalchar}[1]{$^{#1}$}

\appendix

    \section{APS versus MMS for equal entitlements}

\begin{proposition}
    \label{pro:APSMMSXOS}
    For every integer $n \ge 3$ and entitlement $\frac{1}{n}$, there are XOS valuations for which the APS is twice as large as the MMS. 
\end{proposition}

\begin{proof}
    Fix $n \ge 3$ and $m = n + 3$. The valuation is such that there are $n-3$ {large} items that each has value~2. The six remaining items can be thought of as vertices of a graph $G$ that is composed of two disjoint triangles. The value of a non-empty set of vertices is~1, unless its induced subgraph contains at least one edge, and then its value is~2. The APS is~2 (each large item makes a bundle of weight $\frac{1}{n}$, and each pair of vertices that form an edge makes a bundle of weight $\frac{1}{2n}$), whereas the MMS is~1 (because $G$ does not have a perfect matching). 
\end{proof}

\begin{proposition}
    \label{pro:APSMMSSubadditive}
    For every integer $n \ge 2$ and entitlement $\frac{1}{n}$, for every subadditive valuation it holds that $MMS \ge \frac{1}{5}APS$. 
\end{proposition}

\begin{proof}
    Scale the subadditive valuation $v$ so that $APS(\items,v,\frac{1}{n}) = 1$. If there is an item $e$ with $v(e) \ge \frac{1}{5}$,  %(at least $\frac{1}{3}$ would also work), 
    then it can make one of the bundles of the MMS partition. We can then continue by induction on $n$, because $APS(\items \setminus \{e\}, v, \frac{1}{n-1}) \ge APS(\items, v, \frac{1}{n})$. Hence, we may assume that $v(e) < \frac{1}{5}$ for every item $e \in \items$.
    
Consider a fractional APS partition certifying that $APS(\items,v,\frac{1}{n}) = 1$. Because $v$ is subadditive and no item has value above $\frac{1}{5}$, each APS bundle can be broken into two bundles, each of value at least $\frac{2}{5}$. It follows that $APS(\items, v, \frac{1}{2n}) \ge \frac{2}{5}$. This implies that with $2n$ agents that all have valuation $v$, the configuration LP %of~\cite{DNS10} 
has a fractional solution supported only over bundles of value at least $\frac{2}{5}$, in which the contribution of each of the agents to the welfare is at least $\frac{2}{5}$. Rounding this fractional solution using the rounding procedure of Theorem~\ref{thm:roundSubadditive}, each agent has probability at least~$\frac{1}{2}$ of receiving a bundle of value at least $\frac{1}{2} \cdot \frac{2}{5} = \frac{1}{5}$. This implies that there is an integer allocation in which at least $n$ agents get value at least $\frac{1}{5}$. This allocation implies an MMS partition of at least this value.
\end{proof}

\subsection{Computational aspects}
\label{sec:computation}

\begin{definition}
\label{def:capped}
    Given a set function $v$ and a value $C$, we let $v^C$ denote the function satisfying $v^C(S) = \min[v(S), C]$ for every $S \in \items$. A {\em truncated demand query} specifies non-negative prices $\{p_j\}_{j \in \items}$ and a value $C$, and asks for the bundle $S\subset \items$ that maximizes $v^C(S) - \sum_{j\in S} p_j$.  
\end{definition}

\begin{proposition}
\label{pro:computeAPS}
    Let $v$ be a valuation function over $m$ items in which every set has a rational value with numerators and denominators bounded $R$, and let $n$ be a positive integer. Then using  truncated demand query access to $v$, the value of $APS(\items,v,\frac{1}{n})$ as well as a corresponding fractional APS partition can be computed in time polynomial in $m$,$n$ and $\log R$. 
\end{proposition}

\begin{proof}
We sketch the proof. 
%We will find by binary search a value $C$ such that $APS(\items,v,\frac{1}{n}) = C$. For that value, the approach also provides a corresponding fractional APS partition.
Consider a candidate value $C$. Let there be $n$ agents with valuations $v^C$, and consider the configuration LP (CLT). As shown in~\cite{DNS10}, it can be solved in polynomial time using demand queries to the underlying valuation functions $v^C$, which in our case correspond to truncated demand queries to $v$. If the value of the CLT is less than $nC$, this implies that $C > APS(\items,v,\frac{1}{n})$, whereas if the value of the CLT is $nC$, this implies that $C \le APS(\items,v,\frac{1}{n})$. Hence we can use the CLT to do a binary search over the values of $C$. Moreover, for $C = APS(\items,v,\frac{1}{n})$, the CLT solution is a fractional APS partition.   
\end{proof}

The combination of Propositions~\ref{pro:computeAPS} and~\ref{pro:APSMMSSubadditive} implies that if $v$ is subadditive, then with access to truncated demand queries one can compute in expected polynomial time a partition of $\items$ into $n$ parts, such that each part has value at least $\frac{1}{5} MMS(\items, v, \frac{1}{n})$.

\begin{remark}
\label{rem:demand}
    With access to a demand oracle rather than a truncated demand oracle, one can compute in expected polynomial time a partition of $\items$ into $n$ parts, such that each part has value at least $\frac{1}{6} MMS(\items, v, \frac{1}{n})$. We sketch the approach. For $n$ agents with the same valuation function $v$, solve the configuration LP (demand queries suffice for this). Let $T$ be the value of the solution. Necessarily, $T \ge n \cdot MMS(\items, v, \frac{1}{n})$. If there is an item of value at least $\frac{T}{3n}$, allocate it to one of the agents and solve  CLT again with fewer agents. Hence we assume that there are no such items. Use the rounding procedure of {Theorem}~\ref{thm:roundSubadditive} to find (in expected polynomial time) $n$ disjoint bundles of total value at least $\frac{T}{2}$. From each bundle, cut out parts of size between  $\frac{T}{6n}$ and $\frac{T}{3n}$ as long as possible (this uses the assumption that no single item has value above $\frac{T}{3n}$). In each bundle, a value of at most $\frac{T}{6n}$ is left over. Subadditivity implies that this procedure outputs at least $n$ disjoint bundles, each of value at least $\frac{T}{6n}$.
\end{remark}

\section{Proof of {Lemma~\ref{lem:sum}}}
\label{sec:sumProof}

In this section we prove Lemma~\ref{lem:sum}. We first restate it as the following theorem. (For the intended application of Lemma~\ref{lem:sum}, the $x_i$ in the theorem can be assumed to be integers, but the theorem holds also when they are not integers.)

\begin{theorem}\label{sumxbound}
    For integer $k \ge 2$, let $\{x_i\}_{i = -1}^k$, be a sequence of $(k + 2)$ non-negative numbers  satisfying the following properties:
    \begin{enumerate}
    \item $x_{-1} = 0$, $x_0 = 1$ and for all $i \geq 0$, $x_{i - 1} \leq x_i$;
    \item the function $h(i) = x_i - 1$ is \textbf{superadditive}. 
    That is, for all $i, j$, 
    \[h(i + j) \geq h(i) + h(j)\;  \iff \; x_{i + j} - 1 \geq x_i + x_j - 2 \; \iff \;  x_{i + j} \geq x_i + x_j -1;\]
    \item \[\sum_{i = 1}^k\frac{x_{i - 1} - x_{i - 2}}{x_i} < 1.\]
\end{enumerate}
Then $x_k > (k - 1)^{k - 1}$.
\end{theorem}

We define the following variables $y_i = y_i(x)$ for $i \geq 1$:
\[y_i = y_i(x) := \frac{x_{i - 1} - x_{i - 2}}{x_i}.\]

\begin{lemma}\label{pos}
    {Every sequence $x = \{x_i\}_{i = -1}^k$ that satisfies the conditions of Theorem~\ref{sumxbound} is strictly increasing.  Hence, $y_i(x) > 0$ for all $i \ge 1$.}
\end{lemma}

\begin{proof}
{All $y_i$ are non-negative, by property~1 of the sequence $\{x_i\}_{i = -1}^k$. Condition $\sum_{i = 1}^k\frac{x_{i - 1} - x_{i - 2}}{x_i} < 1$ implies that $x_1 > 1$ (otherwise $\sum_{i = 1}^k\frac{x_{i - 1} - x_{i - 2}}{x_i} \geq 1$).
Combining this with superadditivity of $x_i - 1$, for any $i \geq 2$, 
$x_i \geq x_{i - 1} + x_1 - 1$ implies that $x_i > x_{i-1}$.

Finally, observe that $x_{j} > x_{j - 1}$ for all $j \geq 0$ implies that $y_i > 0$ for all $i \geq 1$.}
%y_{i + 1} = \frac{x_i - x_{i - 1}}{x_{i + 1}} \geq \frac{x_{i - 1} + x_1 - 1 - x_{i - 1}}{x_{i + 1}} = \frac{x_1 - 1}{x_{i + 1}} > 0. \]
%Thus, it must hold for all $i \geq 1$ that $y_i > 0$.
\end{proof}

Next, we express each $x_i$ for $i \geq 1$ in terms of $y_1,\ldots, y_k$.
Since $x_{-1} = 0$ and $x_1 = 1$, $y_1 = 1/x_1 > 0$ so $x_1 = 1/y_1$.
Similarly,
\[0 < y_2 = \frac{x_1 - 1}{x_2} \implies x_2 = \frac{x_1 - 1}{y_2} = \frac{\frac{1}{y_1} - 1}{y_2} = \frac{1 - y_1}{y_1y_2},\]
and
\[0< y_3 = \frac{x_2 - x_1}{x_3} \implies x_3 = \frac{x_2 - x_1}{y_3} = \frac{1}{y_3}\left(\frac{1 - y_1}{y_1y_2} - \frac{1}{y_1}\right) = \frac{1 - \sum_{i = 1}^2y_i}{y_1y_2y_3}.\]
In general, for arbitrary $i \geq 3$, we have 
\[0 < y_i = \frac{{x_{i - 1} - x_{i - 2}}}{x_i} \quad \implies x_i = \frac{x_{i - 1} - x_{i - 2}}{y_i}.\]

This allows us reformulate Theorem~\ref{sumxbound} as follows:

\begin{theorem}\label{sumybound}
    Let $\{y_i\}_{i = 1}^k$ be a set of numbers, such that for all $i \geq 1$ we have $y_i > 0$ and $\sum_{i = 1}^ky_i < 1$.
    Define a sequence $\{x_i\}_{i = 1}^k$ as follows:
    \[x_1 = \frac{1}{y_1},\qquad x_2 = \frac{1 - y_1}{y_1y_2}, \qquad \forall i \geq 3 : \; x_i = \frac{x_{i - 1} - x_{i - 2}}{y_i}.\]
    Then, for any such set $\{y_i\}_{i = 1}^k$ it holds that $x_k = x_k(y_1,\ldots, y_k) > (k - 1)^{k - 1}$.
\end{theorem}

First, we are going express $x_k$ only using $y_1,\ldots, y_k$ and no other $x_i$-s for $i < k$.

\newcommand{\Pp}{\mathcal{P}}
\begin{definition}
    Define a directed graph $G^{k}$ as follows.
The vertex set of $G^k$ is $V(G^k) = [k]$, and we draw a directed edge $(i) \to (j)$ if and only if $i \leq j - 2$.
{For $j \ge 0$,} let $\Pp(k, j) = \{(i_0, i_1,\ldots, i_j) :\ \forall s,\ i_s \in [k] \text{ and } i_s \leq i_{s + 1} - 2\}$ denote the set of all directed paths of length $j$ %$j \geq 0$
in $G^k$ {(where the length of a path is the number of edges in it)}.
\end{definition}

\begin{lemma}\label{xkexpress}
    For every $k \geq 2$,
    \[x_k = \left(\prod_{i = 1}^ky_i\right)^{-1}\cdot (1 + Y_k),\quad 
    Y_k = \sum_{j = 0}^{\infty}(-1)^{j + 1}\sum_{\substack{(i_0, i_1,\ldots, i_j) \in \Pp(k - 1, j)}}y_{i_0}y_{i_1}\ldots y_{i_j}.\]
    {\textbf{Note:} by construction, the maximum length of a path in $G^{k - 1}$ is $\lfloor k / 2\rfloor - 1$, so for all $j \geq \lfloor k / 2\rfloor$ the corresponding summands in $Y_k$ are $0$.}
\end{lemma}

\begin{proof}
    We prove the lemma by induction on $k$.
    For the base case, {and so as to give the reader some intuition}, consider $k = 2, 3, 4$.
    As computed earlier,
    \[x_2 = \frac{1 - y_1}{y_1y_2},\]
    and since graph $G^1$  has a single vertex $(1)$, the formula holds.
    Next, {as we have seen,}
    \[x_3 = \frac{1 - \sum_{i = 1}^2y_i}{y_1y_2y_3}.\]
     %\[\ufc{remove? y_3  = \frac{x_2 - x_1}{x_3} \implies} x_3 \ufc{remove? = \frac{x_2 - x_1}{y_3} = \frac{1}{y_3}\left(\frac{1 - y_1}{y_1y_2} - \frac{1}{y_1}\right)} = \frac{1 - \sum_{i = 1}^2y_i}{y_1y_2y_3}.\]
    Since the only paths of length $0$ in $G^2$ are single vertices $(1)$ and $(2)$, we get an expression corresponding to our formula for $x_k$.
    Similarly,
\begin{multline*}
    y_4 = \frac{x_3 - x_2}{x_4} \implies x_4 = \frac{x_3 - x_2}{y_4} = \frac{1}{y_4}\left(\frac{1 - \sum_{i = 1}^2y_i}{y_1y_2y_3} - \frac{1 - y_1}{y_1y_2}\right) \\
    =\frac{1}{y_4}\left(\frac{1 - \sum_{i = 1}^2y_i}{y_1y_2y_3} - \frac{y_3 - y_3y_1}{y_1y_2y_3}\right) = \frac{1 - \sum_{i = 1}^3y_i + y_3y_1}{y_1y_2y_3y_4}.
\end{multline*}
The only paths of length $0$ in $G^3$ are $(1), (2), (3)$, and the only path of length $1$ in $G^3$ is $(1) \to (3)$, since we only allow edges $(i) \to (j)$ if $i \leq j - 2$.
Thus, the expression for $x_k$ holds for $k\leq 4$.

Now, assume that the expression is true up to $x_k$, and we prove it for $x_{k + 1}$.
By the inductive assumption,
\begin{multline*}
    x_{k - 1} = \left(\prod_{i = 1}^{k - 1}y_i\right)^{-1}\cdot (1 + Y_{k - 1}),\\ 
    Y_{k - 1} = \sum_{j = 0}^{\infty}(-1)^{j + 1}\sum_{\substack{(i_0, i_1,\ldots, i_j) \in \Pp(k - 2, j)}}y_{i_0}y_{i_1}\ldots y_{i_j},
\end{multline*}
and
\[x_k = \left(\prod_{i = 1}^ky_i\right)^{-1}\cdot (1 + Y_k), \quad 
    Y_k = \sum_{j = 0}^{\infty}(-1)^{j + 1}\sum_{\substack{(i_0, i_1,\ldots, i_j) \in \Pp(k - 1, j)}}y_{i_0}y_{i_1}\ldots y_{i_j}.\]
By definition,
\[0 < y_{k + 1} := \frac{x_{k} - x_{k - 1}}{x_k} \qquad \implies \qquad x_{k + 1} = \frac{x_k - x_{k - 1}}{y_{k + 1}}.\]
Substituting the expressions for $x_{k}$ and $x_{k - 1}$, we get
\begin{multline*}
    x_{k + 1} = \frac{1}{y_{k + 1}}\left(\left(\prod_{i = 1}^ky_i\right)^{-1}\cdot (1 + Y_k) - \left(\prod_{i = 1}^{k - 1}y_i\right)^{-1}\cdot (1 + Y_{k - 1})\right)\\
    =\left(\prod_{i = 1}^{k + 1}y_i\right)^{-1}\big(1 + Y_k - y_k(1 + Y_{k - 1})\big) \\= \left(\prod_{i = 1}^{k + 1}y_i\right)^{-1}\big(1 + Y_k - y_k- y_kY_{k - 1}\big).
\end{multline*}

We show that $Y_k - y_k- y_kY_{k - 1} = Y_{k + 1}$,
where
\[Y_{k + 1} = \sum_{j = 0}^{\infty}(-1)^{j + 1}\sum_{\substack{(i_0,i_1\ldots, i_j) \in \Pp(k , j)}}y_{i_0}y_{i_1}\ldots y_{i_j}.\]
To see this, we write the definition of $Y_k$ and $Y_{k -1}$.
\begin{multline*}
    Y_k - y_k- y_kY_{k - 1} =\sum_{j = 0}^{\infty}(-1)^{j + 1}\sum_{\substack{(i_0, i_1,\ldots, i_j) \in \Pp(k - 1, j)}}y_{i_0}y_{i_1}\ldots y_{i_j}  \\
    - y_k- y_k\left(\sum_{j = 0}^{\infty}(-1)^{j + 1}\sum_{\substack{(i_0, i_1,\ldots, i_j) \in \Pp(k - 2, j)}}y_{i_0}y_{i_1}\ldots y_{i_j}\right).
\end{multline*}

Consider the graph $G^k$.
We can separate all paths in $G^{(k)}$ into two categories: the ones that include the vertex $(k)$, and the ones that do not include $(k)$.
Then, the only $0$-path containing $(k)$ is the vertex $(k)$ itself, and all the other $0$-paths are exactly paths of length $0$ in $G^{k - 1}$, i.e vertices $(1),\ldots, (k)$.
So,
$\sum_{(i_0) \in \Pp(k, 0)}y_{i_0} = \sum_{(i_0) \in \Pp(k - 1, 0)}y_{i_0} + y_k$,
and in the expression $Y_k - y_k - y_kY_{k - 1}$ both $\sum_{(i_0) \in \Pp(k - 1, 0)}y_{i_0}$ and $y_k$ are multiplied by $(-1)$, which corresponds to the summand $(-1)^{j + 1}\sum_{\substack{(i_0, i_1,\ldots, i_j) \in \Pp(k, j)}}y_{i_0}y_{i_1}\ldots y_{i_j}$ for $j = 0$ in the expression for $Y_{k + 1}$.
Thus, to prove $Y_k - y_k - y_{k}Y_{k - 1} = Y_{k + 1}$, it remains to show that
\begin{multline*}
    \sum_{j = 1}^{\infty}(-1)^{j + 1}\sum_{\substack{(i_0, i_1,\ldots, i_j) \in \Pp(k - 1, j)}}y_{i_0}y_{i_1}\ldots y_{i_j} \\
    - y_k\left(\sum_{j = 0}^{\infty}(-1)^{j + 1}\sum_{\substack{(i_0, i_1,\ldots, i_j) \in \Pp(k - 2, j)}}y_{i_0}y_{i_1}\ldots y_{i_j}\right)\\ = \sum_{j = 1}^{\infty}(-1)^{j + 1}\sum_{\substack{(i_0,i_1\ldots, i_j) \in \Pp(k , j)}}y_{i_0}y_{i_1}\ldots y_{i_j}.
\end{multline*}

Let $j \geq 1$, and consider a path $P \in \Pp(k, j)$ of length $j$ in $G^k$, $P = (i_0, i_1, \ldots, i_j)$.
If $P$ does not contain vertex $(k)$, then $P$ is also a path of length $j$ in $G^{k - 1}$, so $P\in \Pp(k - 1, j)$, and we will have the summand $(-1)^{j + 1}y_{i_0}y_{i_1}\ldots y_{i_j}$ with the correct sign in the remaining expression for $Y_k$.
%By construction, we draw an edge $(s) \to (t)$ in $G^{k - 1}$ if and only if $s \leq t - 2$, hence the maximum length of a path in $G^{k - 1}$ is $\lfloor k / 2\rfloor - 1$, and that is exactly the maximum value of $j$ that we allow in the partial sum for $Y_k$.
%On the other hand, in the sum for $Y_{k + 1}$ we allow values of $j$ to go as high as $\floor{(k + 1) /2} - 1$, maximum length of a path in $G^k$.
%When $k$ is even, $\floor{(k + 1) /2} - 1= \floor{k /2} - 1$, and the maximum length of a path in $G^{k - 1}$ is the same as the maximum length of a  path in $G^k$.
%Hence, every path in $G^k$ that does not contain $(k)$ is also included in $G^{k - 1}$, and is counted in the partial sum for $Y_k$.
%When $k$ is odd, $\floor{(k + 1) /2} - 1 = \floor{k /2}$, so there are paths in $G^k$ longer than any path in $G^{k - 1}$.
%But then, if a path in $G^k$ has length $\floor{(k + 1) /2} - 1 = \floor{k /2}$, it must include vertex $(k)$, and all paths of length smaller than $\floor{k/2}$ without $(k)$ are also paths in $G^{k - 1}$, and we count them in $Y_k$.
So, we obtain an equality
\begin{multline*}
    \sum_{j = 1}^{\infty}(-1)^{j + 1}\sum_{\substack{(i_0, i_1,\ldots, i_j) \in \Pp(k - 1, j)}}y_{i_0}y_{i_1}\ldots y_{i_j} \\= \sum_{j = 1}^{\infty}(-1)^{j + 1}\sum_{\substack{(i_0,i_1\ldots, i_j) \in \Pp(k , j)\\ i_j \neq k}}y_{i_0}y_{i_1}\ldots y_{i_j}.
\end{multline*}

If $P = (i_0, i_1,\ldots, i_j)$ contains vertex $(k)$, then $i_j = k$ and $(i_0, i_1, \ldots, i_{j - 1})$ is a path of length $j - 1$ in $G^k$.
However, by construction we draw an edge $(s) \to (t)$ in $G^k$ if and only if $s \leq t - 2$, hence it must hold that $i_{j - 1} \leq k - 2$.
But then $(i_0, i_1, \ldots, i_{j - 1})$ is a path of length $j - 1$ in $G^{k - 2}$.
Hence, the summand $(-1)^{j + 1}y_{i_0}y_{i_1}\ldots y_{i_{j - 1}}y_{k}$ in the partial sum for $Y_{k + 1}$ can be represented as
\[(-1)^{j + 1}y_{i_0}y_{i_1}\ldots y_{i_{j - 1}}y_{k} = -y_k \cdot (-1)^{j}y_{i_0}y_{i_1}\ldots y_{i_{j - 1}},\]
and
\begin{multline*}
    (-1)^{j + 1}\sum_{\substack{(i_0,i_1\ldots, i_{j - 1}, i_j) \in \Pp(k , j)\\i_j = k}}y_{i_0}y_{i_1}\ldots y_{i_{j - 1}}y_{i_j} 
    \\= -y_k\cdot (-1)^j\sum_{\substack{(i_0,i_1\ldots, i_{j - 1}) \in \Pp(k - 2 , j - 1)}}y_{i_0}y_{i_1}\ldots y_{i_{j - 1}}.
\end{multline*}
%The maximum length of a path in $G^{k - 2}$ is $\floor{(k - 1) / 2} - 1$, while the maximum length of a path in $G^k$ is $\floor{(k + 1) / 2} - 1$.
%When $k$ is even, $\floor{(k + 1) / 2} - 1 = \floor{k / 2} - 1 = (\floor{(k - 1) / 2} + 1) - 1$, hence by considering all paths of length at most $\floor{(k - 1) / 2} - 1$ in $G^{k - 2}$ and adding $(k)$ to them, we will get exactly all paths in $G^k$ of length at most $\floor{(k + 1) / 2}$ that contain $(k)$.
%When $k$ is odd, $\floor{(k + 1) / 2} - 1 = \floor{k / 2} = \floor{(k - 1) / 2} = (\floor{(k - 1) / 2} + 1) - 1$, hence one again we obtain all path of length at most $\floor{(k + 1) / 2} - 1$ in $G^k$ that contain $(k)$ by adding $(k)$ to paths of length at most $\floor{(k - 1) / 2} - 1$ in $G^{k - 2}$.
We conclude that
\begin{multline*}
    \sum_{j = 1}^{\infty}(-1)^{j + 1}\sum_{\substack{(i_0,i_1\ldots, i_j) \in \Pp(k , j)\\ i_j = k}}y_{i_0}y_{i_1}\ldots y_{i_j}\\
    = y_k\sum_{j = 1}^{\infty}(-1)^{j + 1}\sum_{\substack{(i_0,i_1\ldots, i_{j - 1}) \in \Pp(k - 2 , j - 1)}}y_{i_0}y_{i_1}\ldots y_{i_{j - 1}}\\
    = y_k\sum_{j = 0}^{\infty}(-1)^{j}\sum_{\substack{(i_0,i_1\ldots, i_{j}) \in \Pp(k - 2 , j)}}y_{i_0}y_{i_1}\ldots y_{i_j}\\
    = -y_k\sum_{j = 0}^{\infty}(-1)^{j + 1}\sum_{\substack{(i_0,i_1\ldots, i_{j}) \in \Pp(k - 2 , j)}}y_{i_0}y_{i_1}\ldots y_{i_j},
\end{multline*}
and as a result
\begin{multline*}
    \sum_{j = 1}^{\infty}(-1)^{j + 1}\sum_{\substack{(i_0, i_1,\ldots, i_j) \in \Pp(k - 1, j)}}y_{i_0}y_{i_1}\ldots y_{i_j} \\
    - y_k\left(\sum_{j = 0}^{\infty}(-1)^{j + 1}\sum_{\substack{(i_0, i_1,\ldots, i_j) \in \Pp(k - 2, j)}}y_{i_0}y_{i_1}\ldots y_{i_j}\right)\\ = \sum_{j = 1}^{\infty}(-1)^{j + 1}\sum_{\substack{(i_0,i_1\ldots, i_j) \in \Pp(k , j)}}y_{i_0}y_{i_1}\ldots y_{i_j}.
\end{multline*}
We conclude that $Y_k - y_k - y_kY_{k - 1} = Y_{k + 1}$, which proves the inductive step.
Thus, the expression for $x_k$ is correct.
\end{proof}

\begin{lemma}\label{diminish}
    For every $k \geq 2$, for every $j \geq 0$, 
    \[\sum_{\substack{(i_0, i_1,\ldots, i_j) \in \Pp(k - 1, j)}}y_{i_0}y_{i_1}\ldots y_{i_j} \geq \sum_{\substack{(i_0, i_1,\ldots, i_j, i_{j + 1}) \in \Pp(k - 1, j + 1)}}y_{i_0}y_{i_1}\ldots y_{i_j}y_{i_{j + 1}}.\]
\end{lemma}
\newcommand{\Gg}{\mathcal{G}}
\begin{proof}
    Consider the following bipartite graph $\Gg^{k - 1}_j$.
    The vertices $V(\Gg^{k - 1}_j)$ are divided into two disjoint parts labeled by paths in $\Pp(k - 1, j)$ and $\Pp(k - 1, j + 1)$.
    We draw an edge between $S_j \in \Pp(k - 1, j)$ and $T_{j + 1} \in \Pp(k - 1, j + 1)$, where $S_j = (s_0, s_1, \ldots, s_j)$ and $T_{j + 1} = (t_0, t_1, \ldots, t_j, t_{j + 1})$, if and only if $s_0 = t_0, s_1 = t_1, \ldots, s_j = t_j$.
    That is, we have an edge $(S_j, T_{j + 1})$ in $\Gg_j^{k - 1}$, if and only if $T_{j + 1} = (S_j, t_{j + 1})$ for some $t_{j + 1}$.
    Observe that, by construction, $\Gg^{k - 1}_j$ is a bipartite graph that forms a collection of stars centered at paths $S_j \in \Pp(k - 1, j)$. 
    Furthermore, every vertex $T_{j + 1} \in \Pp(k - 1, j + 1)$ is connected to exactly one vertex $S_j \in \Pp(k - 1, j)$, and if $T_{j + 1} = (S_j, t_{j + 1})$ for $S_j = (s_0, s_1, \ldots, s_j)$, then $t_{j + 1} \geq s_j + 2$.
    For a length-$l$ path $P_{l} = (p_0, p_1, \ldots, p_{l})$, denote $y(P_l) := y_{p_0}y_{p_1}\ldots y_{p_l}$.
    Then, we can write
    \begin{multline*}
        \sum_{\substack{(i_0, i_1,\ldots, i_j, i_{j + 1}) \in \Pp(k - 1, j + 1)}}y_{i_0}y_{i_1}\ldots y_{i_j}y_{i_{j + 1}} = \sum_{\substack{T_{j + 1} \in \Pp(k - 1, j + 1)}}y(T_{j + 1})\\
        =\sum_{\substack{T_{j + 1} \in \Pp(k - 1, j + 1)\\ T_{j + 1} = (S_j, t_{j + 1})\\ S_j \in \Pp(k - 1, j)}}y(S_{j})y_{t_{j + 1}} \leq \sum_{\substack{S_{j} \in \Pp(k - 1, j)\\ S_j = (s_0, s_1, \ldots, s_j)}}y(S_{j})\sum_{t_{j + 1} = s_j + 2}^{k - 1}y_{t_{j + 1}}\\
        \leq \sum_{\substack{S_{j} \in \Pp(k - 1, j)\\ S_j = (s_0, s_1, \ldots, s_j)}}y(S_{j}) \cdot \left(1 - y_k - \sum_{t = 1}^{2j}y_t\right)\\
        =  \left(1 - y_k - \sum_{t = 1}^{2j}y_t\right)\sum_{\substack{(i_0, i_1,\ldots, i_j) \in \Pp(k - 1, j)}}y_{i_0}y_{i_1}\ldots y_{i_j} .
    \end{multline*}
    
    The last inequality holds due to the fact that, since $S_j \in \Pp(k - 1, j)$ is a path of length $j$, the index of a vertex $t_{j + 1}$ connected to $s_j$ in a path of length $j + 1$ must be at least $2j + 1$.
    Then
    \[\sum_{t_{j + 1} = s_j + 2}^{k - 1}y_{t_{j + 1}} \leq \sum_{i = 1}^{k}y_i - y_k - \sum_{t = 1}^{2j}y_t < 1 - y_k - \sum_{t = 1}^{2j}y_t,\]
    where $\sum_{i = 1}^{k}y_i < 1$ by assumption.
    We do not make use of the value $1 - y_k - \sum_{t = 1}^{2j}y_t < 1$, hence we just write
    \[\sum_{\substack{(i_0, i_1,\ldots, i_j, i_{j + 1}) \in \Pp(k - 1, j + 1)}}y_{i_0}y_{i_1}\ldots y_{i_j}y_{i_{j + 1}} \leq \sum_{\substack{(i_0, i_1,\ldots, i_j) \in \Pp(k - 1, j)}}y_{i_0}y_{i_1}\ldots y_{i_j} .\]
    {Note that for large $j$ both sides of inequality are $0$, and whenever the right hand side is positive, the inequality is strict.}
\end{proof}

\begin{lemma}\label{ykbound}
    For any $k \geq 2$, $1 + Y_k > y_k$.
\end{lemma}

\begin{proof}
For $j \geq 0$, denote
\[y(k - 1, j) := \sum_{\substack{(i_0, i_1,\ldots, i_j) \in \Pp(k - 1, j)}}y_{i_0}y_{i_1}\ldots y_{i_j},\]
and note that $y(k - 1, j) \geq 0$.
Then, 
\[-Y_k =  -\sum_{j = 0}^{\infty}(-1)^{j + 1}\sum_{\substack{(i_0, i_1,\ldots, i_j) \in \Pp(k - 1, j)}}y_{i_0}y_{i_1}\ldots y_{i_j} = \sum_{j = 0}^{\infty}(-1)^{j}y(k - 1, j),\]
and by Lemma~\ref{diminish}, $y(k - 1, j + 1) < y(k - 1, j)$ for all $j \geq 0$ with $y(k - 1, j) > 0$.
But then, since $-Y_k$ is an alternating sum with terms decreasing in absolute value and a positive first term,
\[-Y_k = \sum_{j = 0}^{\infty}(-1)^{j}y(k - 1, j) < y(k - 1, 0) = \sum_{i = 1}^{k - 1}y_i < 1- y_k,\]
implying that $Y_k > y_k - 1$ and $1 + Y_k > y_k$.
\end{proof}

\begin{proof}[Proof of Theorem~\ref{sumybound}]
    As established in Lemma~\ref{xkexpress}, for every $k \geq 2$,
    \[x_k = \left(\prod_{i = 1}^ky_i\right)^{-1}\cdot (1 + Y_k), \quad
    Y_k = \sum_{j = 0}^{\infty}(-1)^{j + 1}\sum_{\substack{(i_0, i_1,\ldots, i_j) \in \Pp(k - 1, j)}}y_{i_0}y_{i_1}\ldots y_{i_j}.\]
    By Lemma~\ref{ykbound}, $1 + Y_k > y_k$, hence
    \[x_k = \left(\prod_{i = 1}^ky_i\right)^{-1}\cdot (1 + Y_k) > \left(\prod_{i = 1}^{k - 1}y_i\right)^{-1}.\]
    Now, by the inequality between arithmetic and geometric means,
    \[\left(\prod_{i = 1}^{k - 1}y_i\right)^{1/(k - 1)} \leq \frac{1}{k- 1}\sum_{i = 1}^{k - 1}y_i < \frac{1}{k - 1}, \]
    where we used the fact that $\sum_{i = 1}^{k-1}y_i < 1 - y_k < 1$.
    It immediately follows that
    \[\left(\prod_{i = 1}^{k - 1}y_i\right)  < \left(\frac{1}{k - 1}\right)^{k - 1}\]
    and 
    \[x_k > \left(\prod_{i = 1}^{k - 1}y_i\right)^{-1} > (k - 1)^{k  -1}.\]
\end{proof}

\begin{remark}
\label{rem:tighter}
Theorem~\ref{sumxbound} is used in order to deduce that if the number of items satisfies $m \le (k-1)^{k-1}$ for some integer $k$, then the one shot bidding strategy guarantees at least {$\frac{1}{k}$-APS} to subadditive bidders. We note that the implication also works if $m > (k-1)^{k-1}$, as long as no bundle in the support of the fractional APS partition of the agent has more than $(k-1)^{k-1}$ items. Moreover, for an agent with entitlement $b_i$, the proof of the theorem extends to all $m \le \left(\frac{k-1}{1 - b_i}\right)^{k-1}$. 
\end{remark}

\end{document}